\documentclass[conference]{IEEEtran}

\usepackage{amsmath,amssymb}
\usepackage{graphicx}
\usepackage{hyperref}
\usepackage{orcidlink}

\usepackage{booktabs}
\usepackage{xcolor}
\usepackage{xspace}
\usepackage{ragged2e}
\usepackage{caption}
\usepackage{tabularx} 
\usepackage{amsmath}
\usepackage{amsfonts}
\usepackage{listings}
\usepackage{pifont}
\usepackage{arydshln}
\usepackage{microtype}

\title{Trajeform: Embedding Crowd Data}

\author{
\IEEEauthorblockN{
Marilena Lemonari\,\orcidlink{0000-0001-9854-3369},
Panayiotis Charalambous\,\orcidlink{0000-0002-7230-5132},
Yiorgos Chrysanthou\,\orcidlink{0000-0001-5136-8890}
}
}

\begin{document}

\maketitle

\begin{abstract}
  Crowd simulation aims to generate agent behaviours that are rich, natural, and appropriate to their surrounding environment. Recent advances in Machine Learning have enabled such behavioural patterns to be learned directly from trajectory data, allowing simulation models to capture increasingly complex crowd dynamics. However, the capabilities of these systems remain strongly dependent on the type and diversity of the data used during training. When evaluating how well crowd simulators generalise to real-world datasets, most existing metrics focus primarily on trajectory-level differences rather than behaviour-level similarity. In this paper, we present Trajeform, a framework that trains a reconstruction VAE (Variational Autoencoder) on real trajectory data to learn a meaningful continuous embedding that preserves behavioural characteristics of real crowds. The trained encoder can then be applied to synthetic trajectories generated by simulators and used to compare the resulting embedding distributions. Additionally, the learned embedding can be used to analyse supplementary data, such as trajectories corresponding to a simulator’s principal behaviours or captured data from controlled environments (e.g., cycling), providing insight into their ability to span the real behavioural embedding space. The framework compares data from three simulators (CCP, modCCP, and RVO) against ground-truth datasets and is evaluated quantitatively and qualitatively using latent-space analysis, distributional metrics, and trajectory generation through sampling from the aggregated posterior distribution. The results indicate that the learned latent space captures meaningful trajectory dependencies found in the real data, making the framework useful both as an additional measure of synthetic data believability and as a tool for informing the training of future trajectory models.
\end{abstract}  

\section{Introduction}\label{sec:intro}

Crowd simulation aims to populate virtual environments with realistic agent movements. Within this scope, the umbrella of realism includes behaviour diversity, richness, and contextual awareness, in that agents should not only move naturally, but also behave in ways that are appropriate to the environment, e.g., a university campus.
Extensive research has achieved advanced simulators for generating believable agents~\cite{Panayiotou;22, vanToll;20, RVO;08}, utilised in crowd authoring to optimise simulator parameter spaces with respect to an authoring constraint~\cite{MPACT, Wolinski;14}.  
However, most of these systems either rely on manual parameterisation~\cite{Panayiotou;22}, or synthetic data~\cite{MPACT}.
Works that utilise real data are often constrained by the amount and type of data they use (single policy~\cite{Charalambous;23}, controlled environments~\cite{Ju;10, Lee;07}), hence hindering generalisability.
Therefore, it is imperative to measure behavioural similarity by comparing differently sourced agent behaviours with real data~\cite{pathPatterns}.
In this work, we aim to quantify and formalise an embedding representation to assess the quality of agent behaviours in comparison with real ones.
Such a tool enables comparing real and simulated behaviours in the representation space, and informing a selection of parameters spanning the real data features. 

Measuring behaviour similarity properly, in the context of crowds, is a challenging undertaking due to the conceptual definition of a behaviour ``match'' and its non-deterministic nature. 
Relevant works weigh hand-crafted salient features~\cite{Charalambous;14, Wolinski;14}, perceptually validate realism~\cite{mcdonnell2009eye, Hoyet;16}, or combine both (perceptually-validated metric)~\cite{Cabrero;21}. However, these works still focus on \textit{trajectory} faithfulness, and are biased either on the selection of trajectory features, or the expertise of user study participants. Other works utilise entropy-based measures~\cite{Guy;12,Karamouzas;18}. To compare simulated and real data, several systems~\cite{panayiotou2025cedrl, MPACT} report metrics on trajectory features such as Deviation From Goal (DFG), and Optimal Path (OP)~\cite{dijkstra1959note}. Even though these metrics are widely used and can give strong indications regarding the distribution of trajectory features in the two groups (generated and real), they are more tailored to trajectory features in an aggregated manner (e.g., DFG for all agents at every timestep); these partially capture behavioural aspects (turning left and turning right contain the same behaviour but lead to different paths) or limit controllability. We aim to investigate learnable features to complement the assembly of metrics used for comparing between two sets of behaviours.

In this paper, we introduce Trajeform, a trajectory-based framework that learns an embedding space directly from real crowd trajectory data. The learned latent distribution serves as a baseline, against which synthetic trajectories (e.g., generated by crowd simulators) can be compared. Synthetic trajectories are projected onto the learned latent space, allowing their inferred distributions to be juxtaposed with the real-data embedding distribution. The framework preserves the properties of a VAE latent space while encouraging behaviourally similar trajectories to cluster together. Since the model is trained through trajectory reconstruction, the decoder can additionally be used to project latent embeddings back into the trajectory space.
Trajeform primarily aims to project crowd data onto an embedding space tailored to real behavioural features. To achieve this, the framework operates on canonicalised trajectories aligned during preprocessing, encouraging the model to focus on behavioural characteristics beyond simple positional attributes such as spawn location or orientation. Consequently, the learned embedding space captures more behaviour-relevant representations. Beyond measuring behaviour similarity, the framework can also provide indications regarding how effectively a simulator spans the real-data behavioural manifold, while additionally informing simulator parameter selections towards distributions more consistent with the arrangement of real data in the latent space. We evaluate the proposed framework through comparisons with established quantitative metrics, complemented by graphical analyses of the learned embedding distributions.

In a nutshell, we deliver a model trained on real data, that transforms trajectories to learned latent features. The trained latents compile a distribution used as a baseline for comparisons with differently-sourced data, e.g., simulated. Hence, our contribution can be summarised as follows:
\begin{itemize}
    \item Learning to embed crowd data in a feature space aligned with real behavioural characteristics, and
    \item Inferring a latent distribution from real data embeddings, enabling informed sampling, and similarity measurement in latent space.
\end{itemize}
 \section{Related Work}\label{sec:rw_traj}
%

\subsection{Trajectory-based Metrics}\label{subs:rw_metrics}
To compare simulations—either with synthetic or real data—various quantitative metrics have been proposed. At the \textit{macroscopic} level, the Fundamental Diagram~\cite{FD} is widely used, relating speed, density, and flow to capture global crowd phenomena. At the \textit{microscopic} level, metrics focus on per-trajectory features such as deviation from goals, path optimality, and distance to nearest neighbours, which can indicate behaviours like grouping. Another approach compares distributions of trajectory features, such as speed and position.
Several crowd simulators leverage fundamental diagrams and low-level features for validation~\cite{panayiotou2025cedrl, Panayiotou;22, Charalambous;23}.
However, such metrics are more tailored to trajectory replication instead of behaviour similarity and so they are often sensitive to outliers or detailed motions~\cite{huang2019microscopic, silva;24}.
Some works benchmark evaluations, but specifically for steering behaviours~\cite{Singh;09}.
Charalambous et al.~\cite{Charalambous;14DD} propose a method for outlier detection through a data-driven approach, while Lerner et al.~\cite{Lerner;09b} perform database matching. 
More recently, Wang et al.~\cite{pathPatterns} employ Bayesian inference to discover latent path patterns from trajectory data for the analysis and comparison of crowd behaviours.
Our method is complementary to other techniques, and does not perform inference on the trajectory space, but on a compressed latent representation using an end-to-end architecture. 

Perhaps closer to our method is Crowd Space~\cite{Karamouzas;18} that defines agent descriptors (Minimal Predicted Distance) and trains a dimensionality reduction model to find the most suitable simulator for a given real scenario. This method learns a manifold based on the proposed agent-centric entropy~\cite{Karamouzas;18, Guy;12}. In our method however, we do not specify the evaluation metric, but instead, allow the model to learn features freely, and derive a metric post-training, using our learned manifold. Also, we employ a \textit{variational} model, improving the sampling and generative capabilities. Crowd Space aims to assess the predictive accuracy of simulators to find the best simulation method to use on new crowd scenarios. Our method focuses on learning meaningful embeddings from real data, emphasising representation rather than classification. Crowd Space in trained on a larger set of real data that includes controlled and dense environments~\cite{Ju;10, seyfried2009new}, whereas we aim at in-the-wild  behaviours. Finally, in Trajeform, we do not group our trajectories according to their scenario source, e.g., University Campus, but rather, learn a collective representation since behaviours in two different scenarios could be the same. 

\subsection{Perceptual Evaluation}\label{subs:rw_percept}
Apart from using quantitative methods to evaluate trajectory and behaviour plausibility, perceptual evaluation is also explored as a way to mitigate for the aforementioned metric limitations. 
User studies have been implemented to assess trajectory realism~\cite{peters2008crowds, ennis2008perceptual, itatani2024social, giunchi2021perceived}. O'Connor et al.~\cite{oconnor;15} conduct a study to determine whether users find crowd behaviours to be more realistic with a Social Force model, and its implications in psychophysical experimentation. Smaller user studies have also been conducted as part of a larger evaluation protocol of proposed optimisation models~\cite{MPACT}, or authoring tools~\cite{Colas;22} asking users to report on believability and usability. Still, user studies, especially in the domain of virtual crowds, are highly dependent on the study design, the experience of participants in the domain, and general visualisation aspect of the shown stimuli, e.g., body animation, clothing, facial animation, and environment quality~\cite{Yang;20, dang2024hypedsim}.  
Daniel et al.~\cite{Cabrero;21} propose a quality metric that combines comparison metrics with perceptual tools. 
Although existing quantitative and qualitative evaluation approaches have been proven useful in uncovering certain behavioural aspects of simulated agents, in our method we compare directly with real data, using learned features and so alleviating the need to hand-craft quantitative metrics but also avoid the perceptual biases. Our method can be used as an additional approach, complementary to other metrics.

\subsection{Trajectory Representation}\label{subs:rw_embeddings}
Representing trajectories and trajectory-based behaviours is also important for analysis and comparisons. 
Morphable Crowds by Ju et al.~\cite{Ju;10} employs a data-driven method to build a trajectory model and create behaviour prefabs, while
GREIL by Charalambous et al.~\cite{Charalambous;23} learns an inverse Reinforcement Learning policy from real data.
Pure trajectory information usually has higher dimensionality $p=(t,x,z) \in (T,2)$ and contains low-interpretability features in terms of behavioural characteristics, e.g., assigning behaviour labels to a distribution of positional data is not straightforward. Therefore, multiple works have attempted to transform pure trajectories to some compressed representations, e.g., trajectory images~\cite{MPACT, CrowdImprint_}, trajectory parameters~\cite{Wolinski;14}.
Additional works employ learning-based methods to encode trajectory information into high-level parameters, e.g., combination of common behaviours~\cite{Panayiotou;22}, and OCEAN personality attributes~\cite{Durupinar;11}. 
High-level trajectory embeddings enhance explainability and assist comparison and analysis, however, the extracted features are dependent on the data used, which is often not sourced from real scenarios.
Conversely, approaches based on real data, lack controllability and generalisability~\cite{Qiao;19}, especially if said data come from controlled settings instead of in-the-wild. 
Here, we strive for an embedding based on real, in-the-wild data, employing a learning-based method to allow the model freedom to learn appropriate behavioural and trajectory features that would be challenging to define through traditional feature specification.

\subsection{Crowd Data}\label{subs:rw_data}
Having crowd data is essential for improving simulation systems, as it supports data-driven methods~\cite{Lerner;09b} and provides training data for ML models. While both synthetic and real data can be used, real datasets are preferable because they capture complex, hard-to-replicate crowd dynamics. Such data can be used for training (if sufficiently large) or for evaluating new simulators~\cite{Amirian;20, Karamouzas;18}.
Crowd data typically consists of agent trajectories, including at least 2D positions and often orientations. Ideally, datasets would be paired, containing ground truth annotations (e.g., behavioural descriptions), which are crucial for supervised learning. However, collecting such data is challenging. Trajectories are usually obtained either through video tracking or motion capture, with data gathered in controlled settings~\cite{Ju;10, Lee;07} or in real environments ranging from sparse to dense crowds~\cite{seyfried2009new, lammel2013getting}. Due to the difficulty of large-scale motion capture, most datasets rely on video-based tracking.
Widely used datasets include those from~\cite{Lerner;07, CYChurch}, covering environments such as a church yard, a shopping street, and a university campus, as well as video-based datasets like~\cite{Pellegrini;09}. Despite these efforts, there remains a lack of large paired datasets, limiting the ability of learning-based approaches to extract behaviour patterns. As a result, many works rely on synthetic data tailored to their specific needs.
\section{Overview}\label{sec:traj_methodology}
%

\subsection{Problem Definition}\label{subs:problem}
First, let us define some notations. Let $\mathcal{T} \subseteq \mathbb{R}^{2xT}$ be the trajectory space, where $T$ is the maximum trajectory length, associated with path duration which is measured in number of frames - directly corresponding to duration (s) since timestep is kept fixed.  
Also in the trajectory space, let $\tau^{t}$ be the $2D$ position at time $t$, with $\tau \in \mathcal{T}$ a trajectory point. We define $\tau_{C}$ to be the set of category-specific trajectories, with $C=\{R,S,G\}$ denoting real, synthetic, and generated, respectively. 

Our goal is to train an encoder $E_{\theta}: \mathcal{T} \to \mathcal{Z}$, subject to parameters $\theta$, that creates a mapping from the trajectory space to the latent space $\mathcal{Z} \subseteq  \mathbb{R}^{Z}$ with latent dimensions $Z$, where latent point $z \in \mathcal{Z}$. Given the trained encoder, real data $\tau_{R}$ and simulated trajectories $\tau_{f,h}$, we can obtain $z_{R}, z_{f}$ and $z_{h}$ via $z_{i} = E_{\theta}(\tau_{i})$. 
If there are simulators $f: \mathcal{P}_{f} \to \mathcal{T}$ and $h: \mathcal{P}_{h} \to \mathcal{T}$, where $\mathcal{P}_{i}$ a simulator-specific parameter space, suppose that $f,h$ attempt to simulate the real data. Then, comparing the distributions of $z_{f}$ and $z_{h}$ against $z_{R}$, we can provide an indication of how well they represent the real data distribution. 
If $\tau_{f,h}$ are two different datasets to be used for training a trajectory model, the latent distribution comparisons help indicate how close the training set would be to the existing real data - a measure analogous to class balancing. 
%
We theorise that (i) our latent space meaningfully reflects the structure of real data
, and (ii) the properties of our latent space  (continuous, meaningful, preserving real behavioural features) make our encoder more tailored to the real data. As part of this work, we provide evidence of these claims via two sets of experiments (Section~\ref{sec:trajeform_eval}). 

We aim towards a smooth, continuous latent space, that is generative but also has interpretable structure.
This latent space, achieved based on the trained encoder $E_{\theta}$, can be used for: 
\begin{itemize}
    \item Graphical comparison of simulator data (e.g., CCP) with the real data they attempt to simulate (e.g., commercial street).
    \item Distribution comparison between the encoded simulated and real data in the latent space.
    \item Graphical observation of the coverage of the embedding space of differently-sourced data (e.g., principal behaviours of CCP, or data from a controlled environment).
    \item Selection of training data of simulator-specific parameters.
\end{itemize}

\subsection{Model}\label{subs:trajeform_model}
Our goal is to learn an embedding function subject to trajectory information from real data. Considering the real data size, we incorporate encoder/decoder modules that work well on lower cardinality. Thus, we train a VAE on the task of trajectory reconstruction, and define our learned latent space as the desired manifold. We approach the training setup as follows: (a) we train a VAE on reconstruction without explicitly incorporating a prediction of the generated trajectory length, and (b) learn an additional feed-forward model to predict the path length $[l_{i}; i \leq L_{max}]$, from the latent space $\mathcal{Z}$. We do this since our data vary in path durations and so we help our decoder generate more appropriate paths at inference time, by inputting the predicted length in the form of a mask matrix. Hence, the whole model is trained on the loss given in Equation~\ref{eq:loss}, with $\beta=.1$. The low value for $\beta$ comes after experimentation, trading off reconstruction quality and latent space faithfulness to prior distribution $\mathcal{N}(0,1)$; other works with related concerns use small KL-divergence weight~\cite{shao2020dynamicvae,neumeier2021variational}. 
\begin{equation}
\begin{aligned}
    \mathcal{L} = \mathcal{L}_{VAE} + L_{len.} &\text{ ; }
    \mathcal{L}_{VAE} = \mathcal{L}_{recon.}+\beta \mathcal{L}_{KL} \\
    \mathcal{L}_{recon.} &= \text{MSE}(\tau)\\
    \mathcal{L}_{len.} = \text{CrossEntropy} &\text{ on trajectory len. classification} \\
\end{aligned}
\label{eq:loss}
\end{equation}

\begin{figure}[h]
  \centering
  \includegraphics[width=\linewidth]{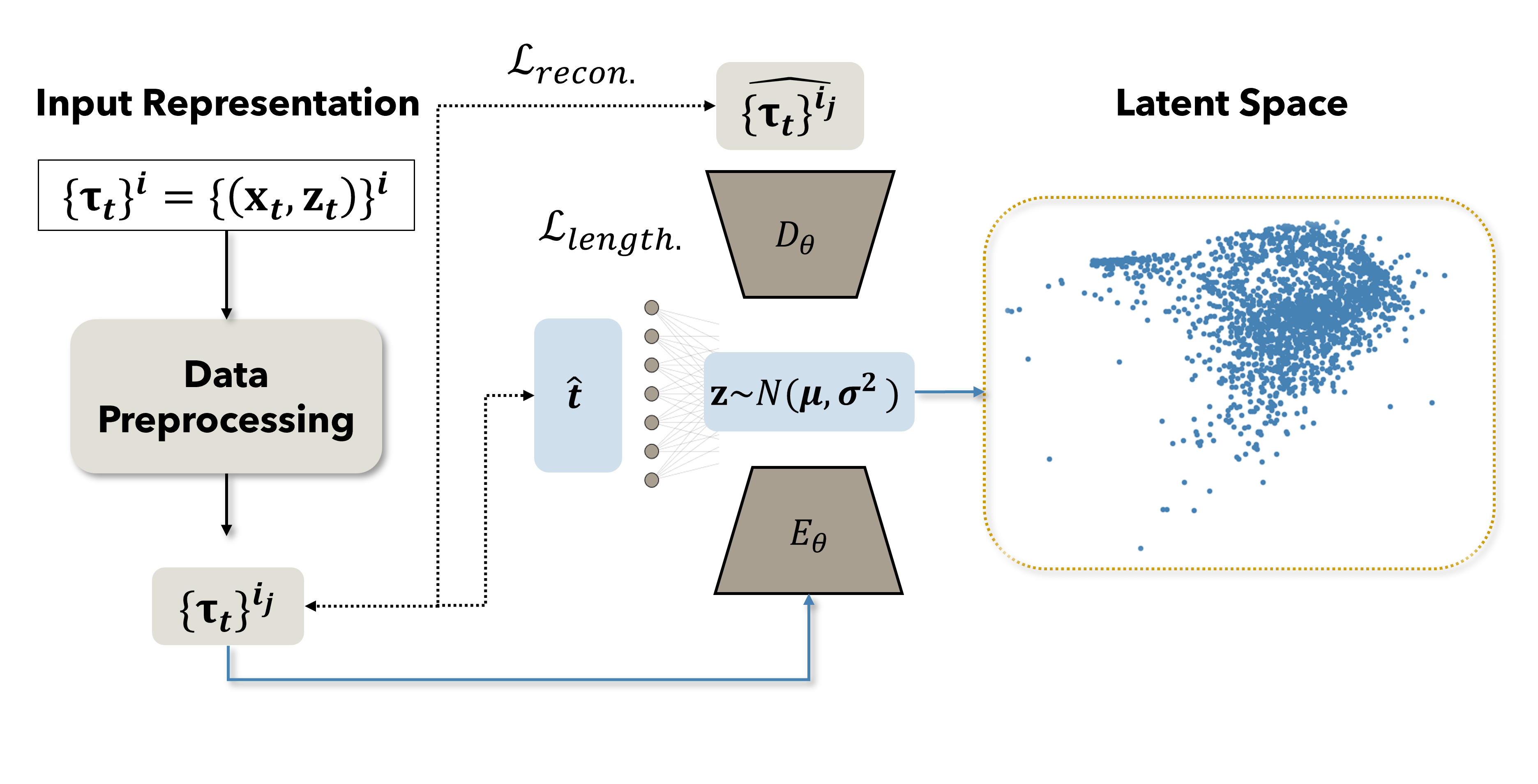}
  \caption{\label{fig:conv2gru} Model architecture.}
\end{figure}

We experiment with the model architecture, and specifically the architecture for the encoder and decoder blocks. Temporal convolutions are good analysers and could prove useful for feature extraction in our case. At the same time GRUs provide smooth, temporally processed generations. Therefore, we conduct a model tuning study where we experiment with block combinations, as shown in Table~\ref{table:modelArchitecture2}. The table reports the latent losses $L_{z} = ||z - \tilde{z} ||_{2}$ on a common test set. 
Note that we do not include transformer blocks due to the small training dataset. In the case of a convolutional decoder, the reconstructions return jerky paths due to the block's focus on preserving local structure, while GRU generators favour smoothness. Since our goal is to create a structured latent space, where similar behaviours in trajectories could be gathered together, we also employed a Mixture of Gaussians as a prior. Ultimately, visualising the quality of the test reconstructions and latent space projections, we proceed with a convolutional (encoder) to GRU (decoder) architecture, denoted conv2gru. Figure~\ref{fig:conv2gru} illustrates the architecture overview.  
%
%
\begin{table}[h]
\centering
\begin{tabular}{c | c} 
 \hline
 Architecture & Test Latent Loss$\downarrow$ \\ [0.5ex] 
 \hline
 conv2gru      & \textbf{.381} \\
 conv2conv     & \textbf{.304} \\
 gru2gru       & .440 \\
 gru2conv      & .414 \\
 conv2gru\_mog & .443 \\
 \hline
\end{tabular}
\caption{Cross architecture experimentation on the complete dataset assembly (Flock, Students, Zara, and ETH)}.
\label{table:modelArchitecture2}
\end{table}

\subsection{Model Data}\label{sec:model_data}
The model has been trained on a set of real data trajectories. In this section, we give details regarding the data processing.

\paragraph*{Real Dataset}\label{subs:realData}

We use existing tracked datasets corresponding to trajectory data captured in a church yard, a university campus, a commercial street, and outside a hotel, henceforth referred to as Flock, Students, Zara, and ETH~\cite{CYChurch, Lerner;07, Pellegrini;09}. 
The total number of trajectories for each dataset is $\{24, 849, 531, 750\}$, respectively, sampled at a timestep $=.4$, for uniformity. 
Before continuing to the preprocessing step, we exclude Zara3 from our experiments dataset, and save it for testing. Finally, the datapoints are padded according to the maximum length, and a relevant mask is produced.

\paragraph*{Data Preprocessing}\label{subs:preprocessing}

\begin{figure}[h]
  \centering
  \includegraphics[width=\linewidth]{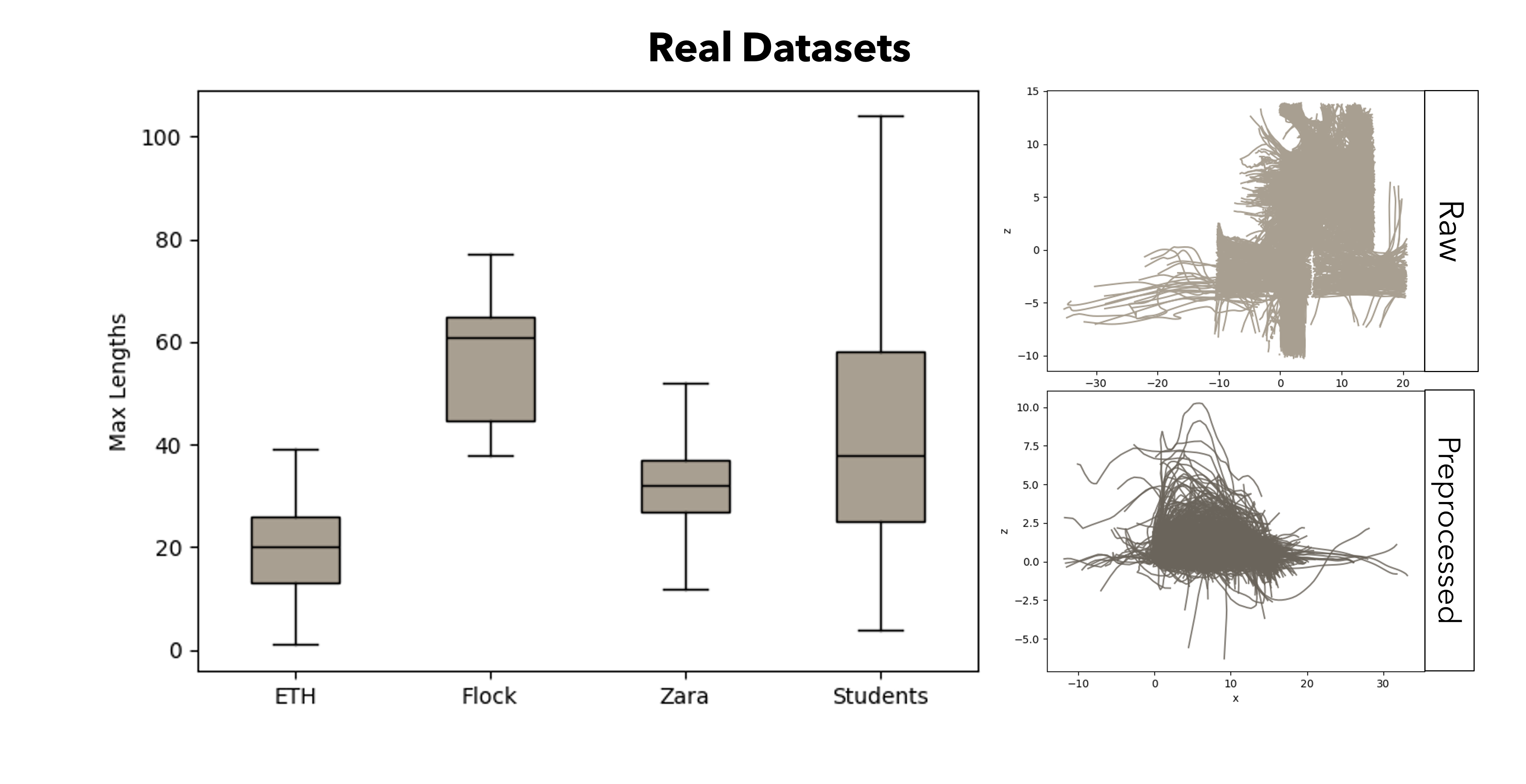}
  \caption[Real data features.]{\label{fig:cumulative}
           Real Data Features. 
           }
\end{figure}

Prior to training the model, we pass the data through a preprocessing module whose purpose is to canonicalise it in order to help the model learn useful features and not be, for example, biased on the starting position.
This module includes: path segmentation, transformations, spatial canonicalisation, and noise addition. More specifically, we disregard paths with $L<L_{min}=5$, and segment paths with $L>L_{max}=104$ to smaller valid parts. The transformations include, translation to the (x,z) origin, rotation of the direction of movement towards the right x-axis, and flipping of x and z-axes to the first quadrant. 
For spatial canonicalisation, we try both bounding the (x,z) space, and normalising according to $(z-z_{mean})/z_{std}$.
Finally, for noise augmentation we randomly add noise $\epsilon$ sampled from a standard normal. 
The choice for $L_{max}$ is justified by the boxplot in Figure~\ref{fig:cumulative} which omits the outlier lengths for each dataset. It also shows an example preprocessing module that includes segmentation and transformation as mentioned. For a more detailed visualisation of the preprocessing components, the reader is referred to the supplementary material.

We ablate the individual components of the preprocessing module. The ablation experiments were conducted with the same underlying model architecture and hyperparameters, for fairness. 
We do not report on the model losses because the scale is not comparable, e.g., in normalisation, lower losses do not imply better model, or without translation, the model achieves better performance because it learns to reconstruct not according to the trajectory/behaviour features but according to spawn.  
The final module 
includes the segmentation and transformation operations, and thus is used in the inference pipeline; note that padding and translation to origin are mandatory. 
The final training set consists of \textasciitilde2k samples, and $maxLen$ of $104$ frames.
%
%

\subsection{Embeddings Visualisation}\label{subs:modeldataAnalysis}

\begin{figure}[h]
  \centering
  \includegraphics[width=\linewidth]{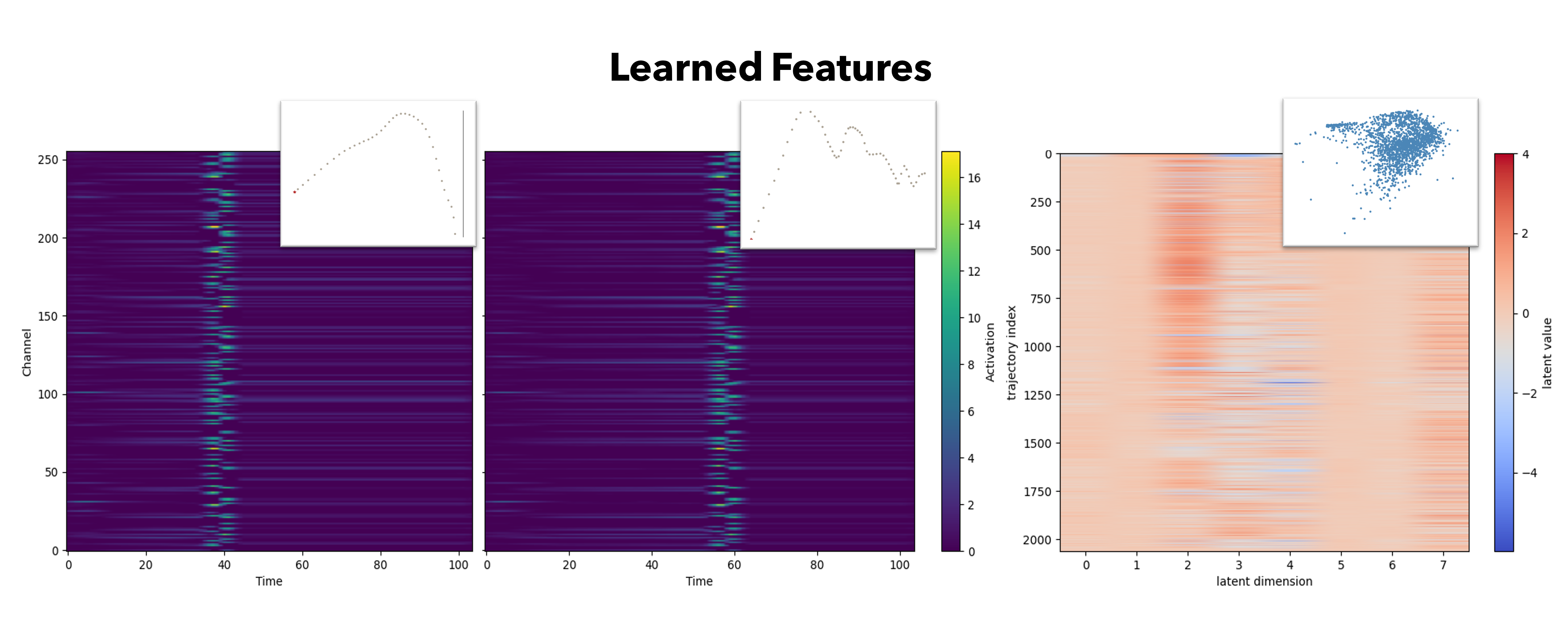}
  \caption[Model feature maps.]{\label{fig:learnedFeatures} Model Feature Maps. Instances of convolution features for two different trajectory samples (left and middle), and the heatmap of the learned latent layer (right).}
\end{figure}

After training the VAE model, we visualise some feature maps, as presented in Figure~\ref{fig:learnedFeatures}. From the figure, we can see that two different trajectory points produce heatmaps of the same convolutional feature that have the same overall structure but differ in the temporal ``activation''. On second inspection, these activation frames correspond to the number of unpadded frames in the respective samples, indicating that this feature learns the path length. In Figure~\ref{fig:learnedFeatures}-right, we can see the heatmap visualisation of the latent layer, revealing the usage of latent dimensions for representation. 

\paragraph*{Implementation Details}
Models have been trained using an 11th Gen Intel(R) Core(TM) i7-11800H @ 2.30GHz, 2304 Mhz, 8 Core(s), 16 Logical Processor(s) CPU, 16GBs of RAM and an NVIDIA GeForce RTX 3070 GPU. Training took approximately 1.5hours, for 200 epochs (100 the VAE, and 100 the length prediction layer), a batch size of 16, dropout of .2, and 8 latent dimensions, with Adam optimizer and learning rate of $3e^{-4}$.
\section{Results}\label{sec:trajeform_results}

At inference time, we utilise the trained model to cluster the real latents, and build a MoG distribution according to the cluster centres and densities. We use this \emph{posterior} to: test on our control data (Zara3), infer principal trajectories from the posterior means, and sample the distribution to generate informed paths, hence crafting a latent similarity metric between inference and training points.

\subsection{Preliminary Results}

\begin{figure}[h]
  \centering
  \includegraphics[width=.9\linewidth]{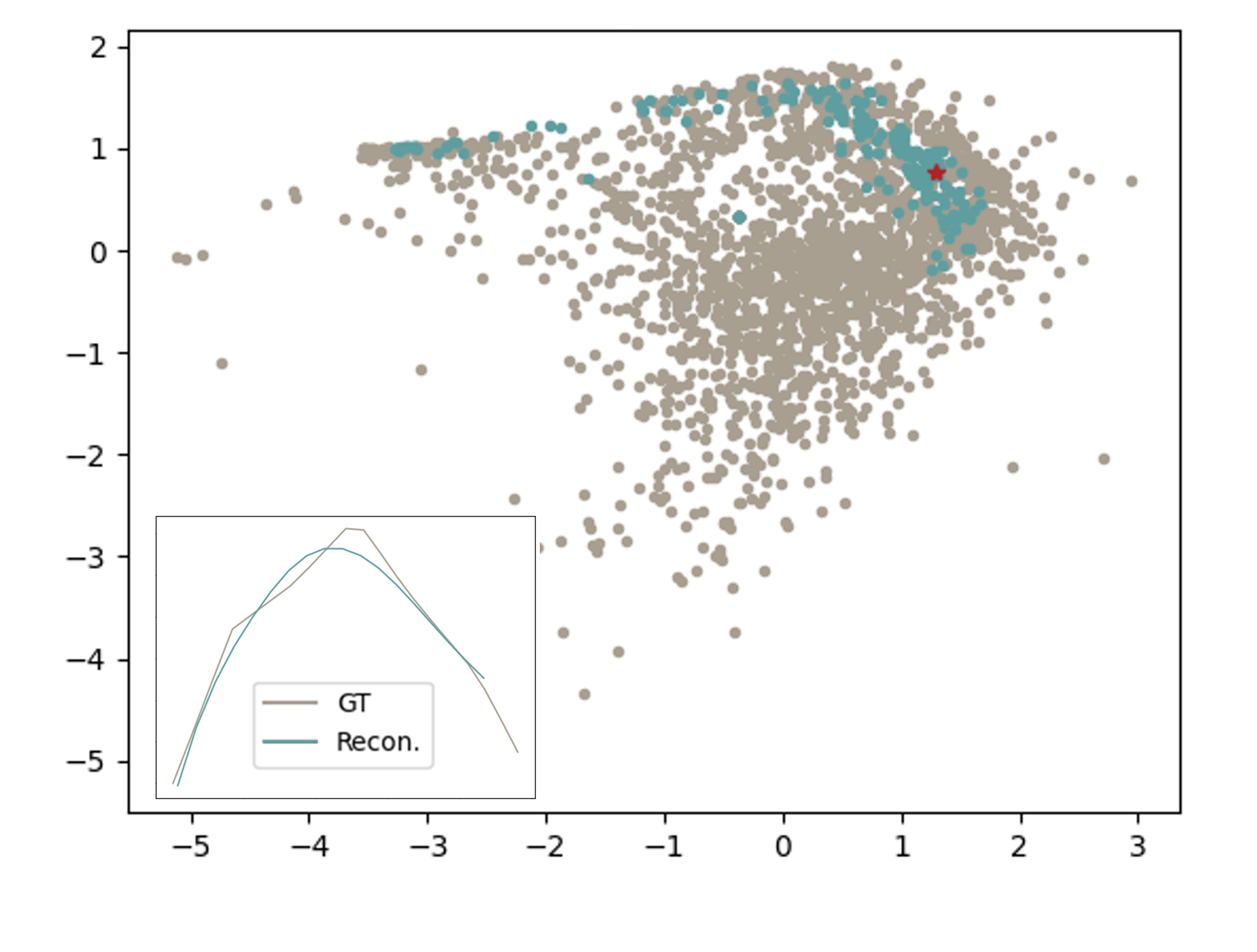}
  \caption{\label{fig:testZara3} Trained and test latent values.}
\end{figure}

Firstly, we take the test dataset we had set aside (Zara3), pass it through the preprocessing block and apply the trained model. The results yield a test reconstruction loss of $.959$ and a length loss of $3.634$. The latent representation of the test trajectories overlay the training latents in Figure~\ref{fig:testZara3}; the red latent's sample trajectories are also shown in the bottom left (original and reconstructed). The 8-dimensional latent space was visualised via PCA dimensionality reduction technique~\cite{pca}. From the visualisation, we can see that the learned latent space encodes the real data in a structured manner, with the test latents fully falling into the trained latent area. We also report the mean latent loss $L_{z} = ||z - \tilde{z} ||_{2} = .582$. This loss was not used during training, but provides an indication of the latent space geometry.  
More results on the model training can be found in the supplementary material. 

For further testing, we re-train the model on four variants, i.e., holding out each of the four datasets (Flock, ETH, Students, Zara). Then, having these ``hold-out models'' we test the model by projecting the “unseen” data (e.g., Flock data on the model variant that has been trained on Students-ETH-Zara). Table~\ref{table:rebuttalTesting} documents the mean latent loss; for the latent space projection of the test data, refer to the supplementary material. 

We compare the four test datasets on the full model and on the respective hold-out models. We observe that when leaving out the Flock dataset, the latent loss actually decreases which makes sense since Flock contains few trajectories. We also see that leaving out ETH has the biggest effect on the latent loss hinting that ETH contains behaviours that are rare in the rest of the datasets (this can also be confirmed from the figure that shows a lots of ETH latent points clustering on the top right of the space). This loss difference can also be attributed to the size of the ETH datasets (2nd largest). The average increase of latent loss from the full model to the hold-out variants is from $.977$ to $1.242$, which shows that our model is robust in that it does not just memorise dataset-specific patterns, but rather, unseen real datasets still map coherently into the learned behavioural space. An increase is however expected since the amount of real data is limited (\~{}2k).
\begin{table}[h]
\centering
\begin{tabular}{c|c c } 
 \hline
  Mean Latent Loss$\downarrow$ & w/ full model & w/hold-out\\ [0.5ex]
 \hline
 w/o Flock & 1.123 & 1.012\\
 w/o Students & .775 & .952 \\
 w/o Zara & .724 & .842\\
 w/o ETH & 1.286 & 2.160\\
 \hline
\end{tabular}
\caption{
Experiments of the effect of holding out a dataset for training, and then testing using the excluded data.}
\label{table:rebuttalTesting}
\end{table}

\subsection{Latent Space}\label{subs:latentSpace_traj}

\begin{figure}[h]
  \centering
  \includegraphics[width=.9\linewidth]{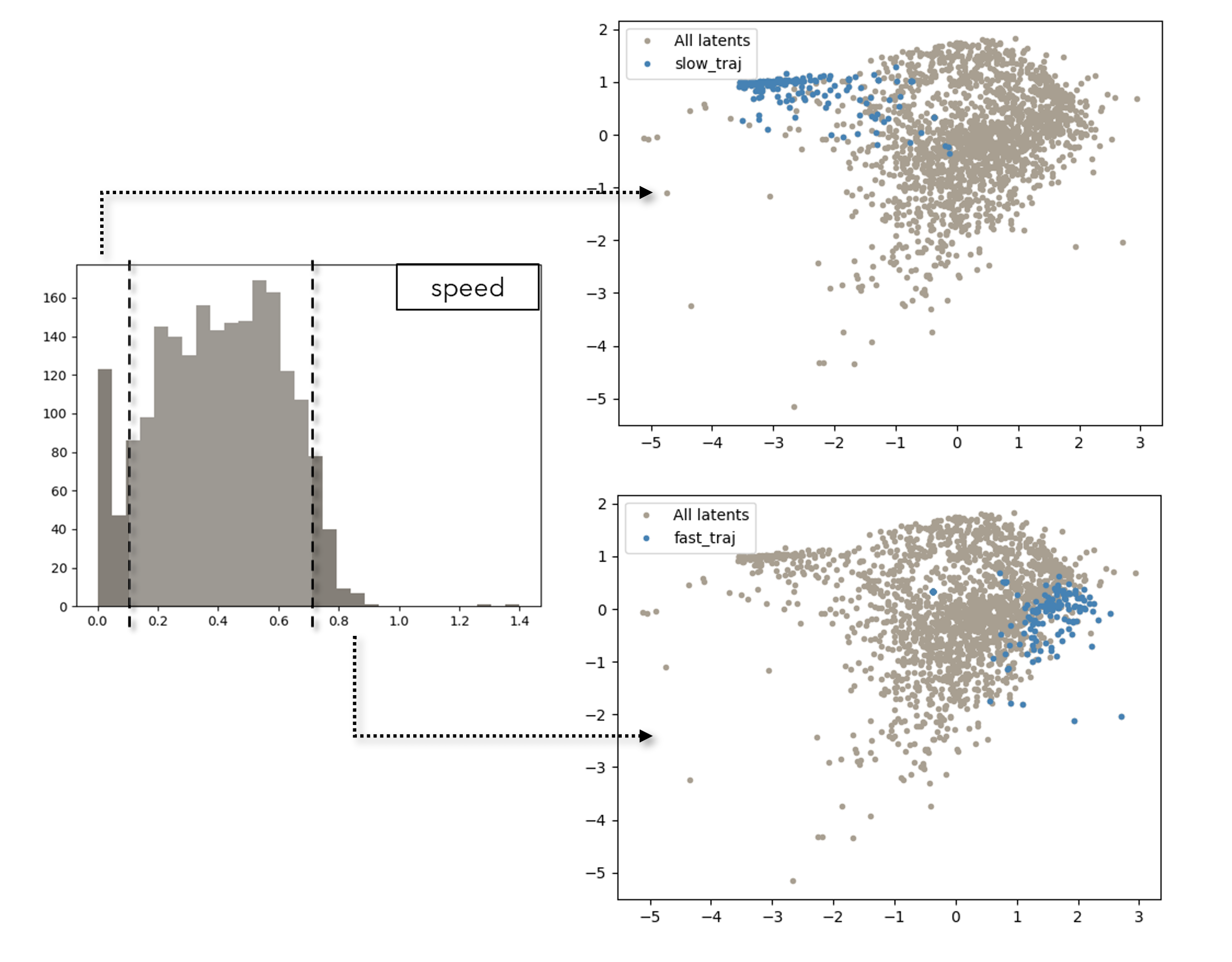}
  \caption{\label{fig:interesting}Latent visualisations of targeted samples.}
\end{figure}

After the preliminary results, we dive deeper into the properties and capabilities of the learned embedding. Firstly, we select families of trajectories in the training data, that we deem ``interesting'', and visualise them to inspect for patterns. Figure~\ref{fig:interesting} for example, shows the distribution of speed value in the datasets (left), along with the plotted interesting groups: in this case, the fast and slow trajectories decided via thresholds $v_{min} = .1$, and $v_{max} = .7$. The results show a clear gathering of the two families in opposite ends of the latent space. We confirm this latent property via also inspecting respective results for long-short trajectory pairs, and for agents with high or low variability in their displacements $(\Delta x, \Delta z)$ during their paths (the reader is referred to the supplementary material for visualisations). 
\begin{figure}[b]
  \centering
  \includegraphics[width=\linewidth]{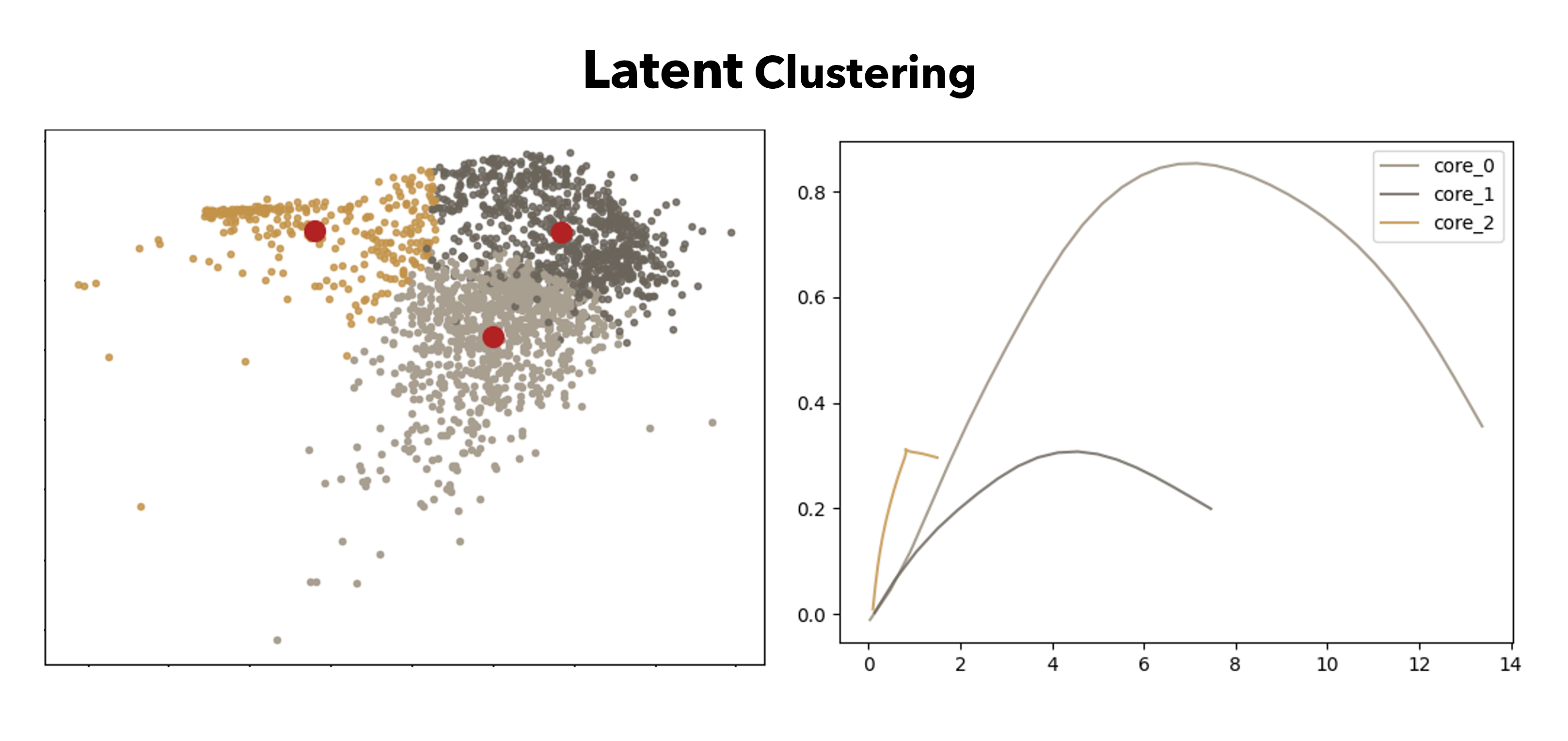}
  \caption{\label{fig:core_traj} Latent clustering and cluster centres' principal behaviours.}
\end{figure}
\begin{figure*}[h]
  \centering
  \includegraphics[width=.9\linewidth]{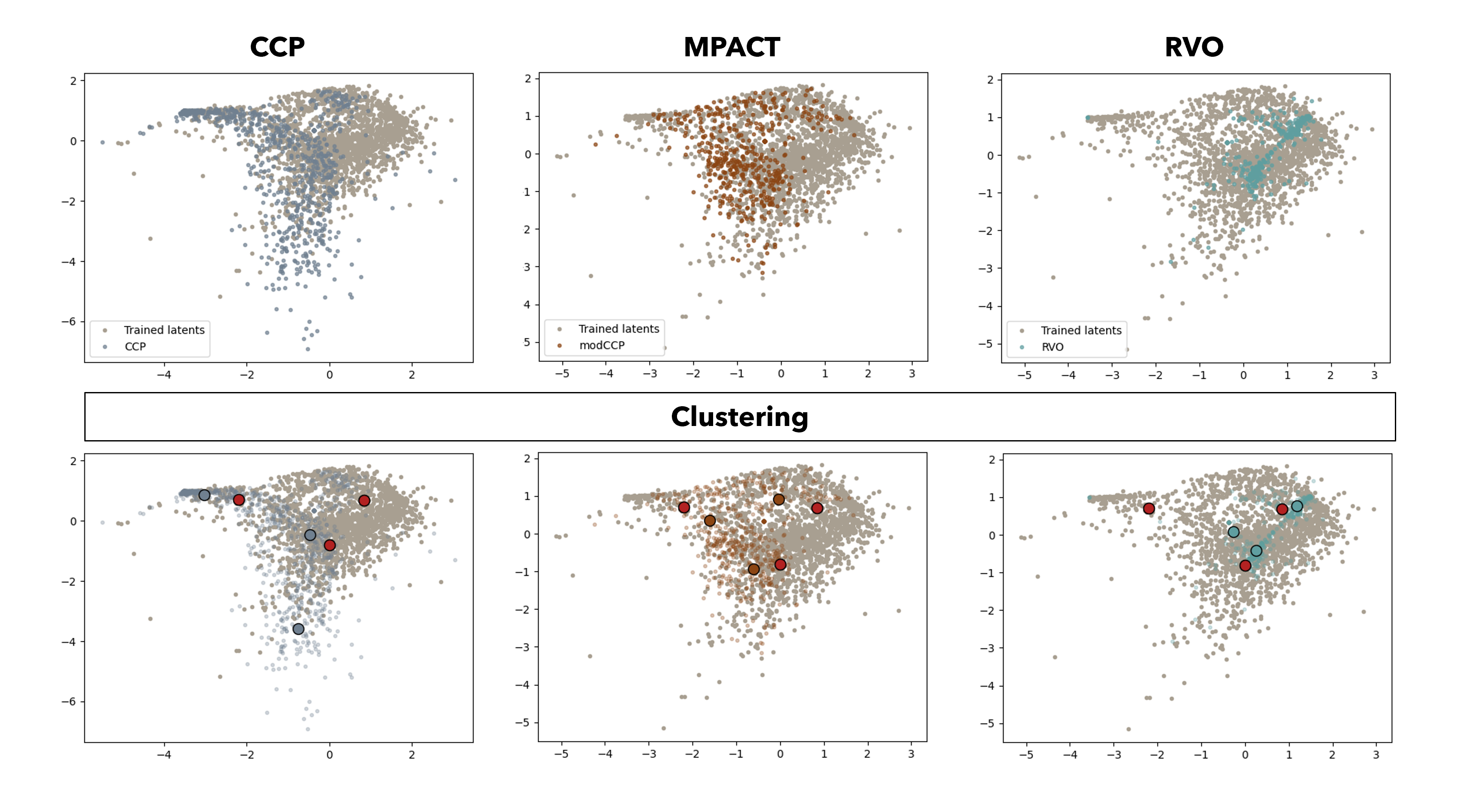}
  \caption{\label{fig:usage_traj} Framework usage on simulator-specific data, mimicking the real data from three simulation types, with their cluster centres.}
\end{figure*}
\begin{figure*}[h]
  \centering
  \includegraphics[width=\linewidth]{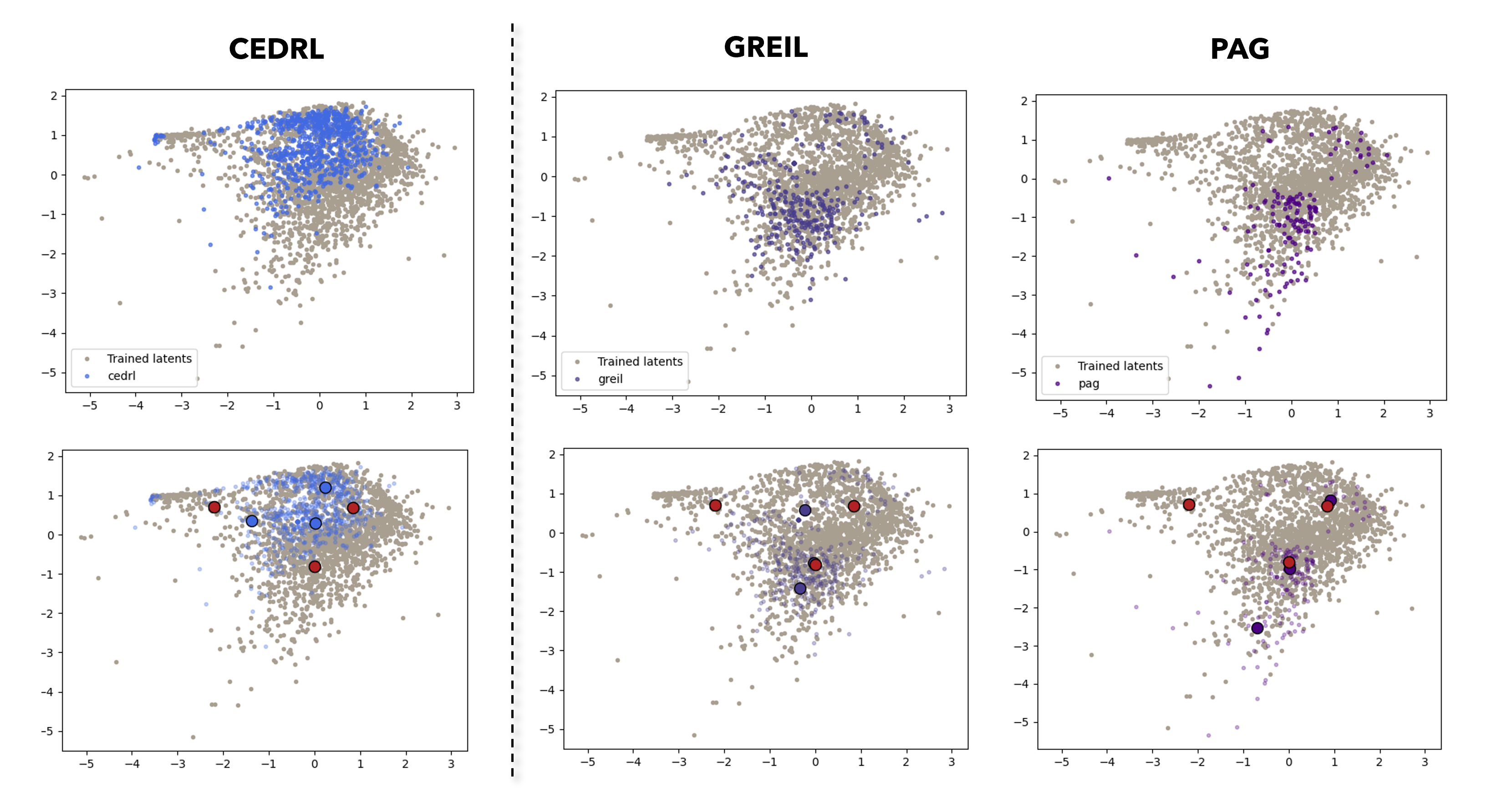}
  \caption{\label{fig:ioannis_exp} 
  Framework usage on additional simulator-specific data, i.e., CEDRL, and the Zara-specific GREIL and PAG-Crowd data.}
\end{figure*}

\begin{table}[b]
\centering
\begin{tabular}{c|c c c} 
 \hline
 Dataset & Recon. Loss$\downarrow$ & Active dims$\uparrow$ & Latent Entropy$\uparrow$\\ [0.5ex]
 \hline
 All & .230 & 7 & .916\\
 \hdashline
 Students & .323 & 7 & .956\\
 Zara & .317 & 8 & .845\\
 ETH & .078 & 4 & .919\\
 \hline
\end{tabular}
\caption{
Testing for VAE model collapse by training dedicated per-dataset models.}
\label{table:ablationDATASET}
\end{table}  
%
VAEs frequently suffer from mode collapse and so to address these concerns, we conduct ablation studies for: training separate spaces per dataset, and comparing against a VQ‑VAE. 
For the former, Table~\ref{table:ablationDATASET} reports (for each separately-trained model variant and the full model) the following metrics: the validation reconstruction loss, the number of active dimensions, and the latent entropy. While training the per-dataset variants, we find that the Flock-only variant overfits - which makes sense since the dataset size is considerably limited. Therefore, we exclude it for further observations. Regarding the reconstruction loss, the full model reaches $.230$ while the average of the three per-dataset variants is $.239$, showing faithful trajectory representation. Regarding the number of active dimensions (in our 8-dimensional latent space), we find based on the reported values that the latent space is indeed used. Lastly, regarding the normalised latent entropy which we compute to assess whether the latent usage is diverse, Table~\ref{table:ablationDATASET} shows high values across variants, and so the representation is not concentrated into a few modes. Collectively, these verify that there is no mode collapse. 
We also train a VQ-VAE variant that essentially replaces the continuous latent variables of a VAE with a finite codebook of learned embeddings, and report a validation reconstruction loss of $1.187$ (larger than our VAE). The loss increase along with the our intention to have a continuous space which we can later use for sampling to map back to the trajectory space (through the decoder), provide evidence for the usefulness of the continuous VAE latent space.

Then, provided a structured and meaningful latent space, we can cluster the embeddings of our training samples, which come from real data. 
This clustering takes place in the latent space and not in the PCA-projected space.
The intuition is that this clustering can be used to build a behaviour distribution tailored to the real data embeddings. We could then use the cluster centres to find the $k$ most common paths, simply by feeding the cluster points into our decoder. Since our model is variational, the latent space is continuous and so this works even if these points do not belong in the original dataset. We use k-means clustering~\cite{kmeans} to find the groups. Figure~\ref{fig:core_traj} shows such clustering along with the corresponding paths. Given a dataset, this process can be used to obtain the ``core behaviours'' found within that set. 
We choose to proceed with k=3 number of latent clusters after both visual observation of the latent space, as well as an ablation analysis (presented and discussed in the supplementary material).
As a reminder the generated paths are in canonical form (translated, rotated, and flipped). Variations of these behaviours can be obtained when distributing in the environment by inverting the transformations with randomised parameters; 
further analysis on intra and inter cluster paths can be found in the supplementary material.
In our real data example, we observe that the cluster centres correspond to paths with substantial differences in distance covered, with the shorter core path corresponding to a sparse cluster; this also confirms the findings in Figure~\ref{fig:interesting}. 

\subsection{Generations}\label{subs:generations}
%
Due to the variational component of our model, we can sample the latent space and generate corresponding trajectories. There are three main ways to sample the space: traversing the space, sampling according to the prior (resembles standard normal), or building some posterior distribution. Our aim is to be able to assign meaning to the distribution of the real behaviours in the latent space. Therefore, having the clustering of Section~\ref{subs:latentSpace_traj}, we can build a MoG with means equal to the cluster centres and mixture weights proportional to the cluster sizes, i.e., $p(z) = \pi_1 \,\mathcal{N}(z \mid \mu_1, \Sigma_1)
      + \pi_2 \,\mathcal{N}(z \mid \mu_2, \Sigma_2)
      + \pi_3 \,\mathcal{N}(z \mid \mu_3, \Sigma_3)$. 
Supplementary material shows instances of sampling 1k points for each method.

We can further operate on the latent space by leveraging both the encoder and decoder modules. 
Specifically, assume that we have a candidate trajectory (e.g., generated by a simulator) that we wish to compare against a real trajectory counterpart. 
The two trajectories are first projected through the trained encoder onto the latent space, producing $z_s$ for the unseen latent (synthetic trajectory) and $z$ for the reference latent (real trajectory). Their latent difference can then be computed as:
\[
||z-z_s||_2.
\]

Since the decoder is trained on real data, we can additionally move from the latent space back to the trajectory space by decoding the reference latent:
\[
\tilde{\tau} = D(z),
\]
and subsequently re-projecting the reconstructed trajectory back into the latent space:
\[
\tilde{z} = E(\tilde{\tau}).
\]

Conceptually, the reconstructed latent $\tilde{z}$ should preserve the same behavioural attributes as the original latent $z$. Therefore, the quantity
\[
||z-\tilde{z}||_2
\]
provides an indication of the reconstruction-consistent latent variation that still corresponds to the same underlying behaviour. 

Consequently, the normalised latent metric
\[
\frac{||z-z_s||_2}{||z-\tilde{z}||_2}
\]
can be used as a measure of behavioural similarity between two encoded trajectories, where values $<1$ indicate a behavioural match.

Table~\ref{table:trajeform_eval2} reports this metric for selected examples, which are discussed in more detail in the following sections.
\section{Use Cases}\label{sec:use_Cases}

In this section we explore three use cases. Mainly, we compare simulator data that mimic the real, with the ground truths through latent distribution comparisons to reveal behaviour similarity. Then, we test other data sources, e.g., principal simulator behaviours or captured data from controlled environments, to inspect how well the span the latent space, and lastly, use the distribution of the real latents to tune simulator-specific parameters in order to guide simulations towards more ``real-data'' balanced generations.

\subsection{Behaviour Similarity}\label{subs:case}

We show a use case of our framework, essentially showing the latents of differently generated synthetic data and comparing to see which is closer to the real data latent MoG.

Firstly, we gather synthetic data from three sources that attempt to mimic the real data behaviours. Specifically RVO, CCP, and MPACT~\cite{RVO;08,Panayiotou;22,MPACT};  
for more details on the respective simulators/methods, please refer to the supplementary material. 
The parameters for RVO are set to match the real data, whereas CCP denotes the behaviour produced when manually setting a single CCP profile that best matches the real data by inspection (both reported in~\cite{MPACT}).
The latter, modCCP, refers to the modCCP behaviour profiles inferred and optimised through the MPACT model. Henceforth, we refer to the optimised modCCP weights as MPACT, for consistency and readability; other references to modCCP imply controlled profiles of the simulator itself - not optimised by the MPACT model.
Therefore, we expect that our metric based on MoG latent distributions, is better for MPACT trajectories.

Each set of simulator-specific trajectories is passed though our pipeline and juxtaposed against the real latents. Then, a new posterior distribution can be defined according to the new clustering, and is compared to the original. Figure~\ref{fig:usage_traj} demonstrates the plotted latents for each inference dataset. Figure~\ref{fig:usage_traj} also shows the visualisations per-simulator, along with the new distribution means.
%
%
From the figures, we can observe that the MPACT latents are closest to the real ones, while RVO values tend to be confined and not access the whole latent space. This could be due to the limited behavioural variability of the simulator itself, since its target use is for generating local movements. We also calculate the metric introduced in Section~\ref{subs:generations} between the pairs of real and simulated cluster centres (Table~\ref{table:trajeform_eval2}). 
The averaged metric value is lower for MPACT-predicted trajectories, confirming our expectations.

The method is generalisable, and so we can input additional simulator-specific data (Figure~\ref{fig:ioannis_exp}) that mimic the real behaviours, specifically from CEDRL~\cite{panayiotou2025cedrl}, GREIL~\cite{Charalambous;23}, and PAG~\cite{Charalambous;14}. We can see that the CEDRL distribution is close to the real ones, yielding a latent metric of $2.006$, placing it second best after MPACT ($1.867$). We can also restrict the data we use to a single scene, e.g., Zara, and plot the Zara-generated data for GREIL and PAG to assess their embeddings when focusing on a specific scenario. Both graphically and according to the latent metric, GREIL outperforms PAG (latent difference $\approx.5$), which makes sense since GREIL is a more recent method, trained on real data with inverse RL and is consistent with observations made in the original  paper (\cite{Charalambous;23}).
GREIL was trained on three datasets, namely ``Chat and Flock-1''~\cite{Ju;10}, and Flock and Zara. For new datasets it requires a separate trained policy.

Overall, users can utilise this framework by directly inputting a set of trajectories which simulates the real scenarios (e.g., church yard, commercial street, university campus) and observe how close they are to the real data features, qualitatively, by visualising the posterior distributions and cluster means in the learned embedding space, or quantitatively, by calculating a latent metric. Users can also sample the learned space to inspect generated trajectories. 

\subsection{Embedding Space Coverage}\label{subs:case2}
We can also use the model on data that do not simulate the real scenarios. In that case, the model is used as a way to visualise how spread the inputs are in the embedding space.

\begin{figure*}[h]
  \centering
  \includegraphics[width=\linewidth]{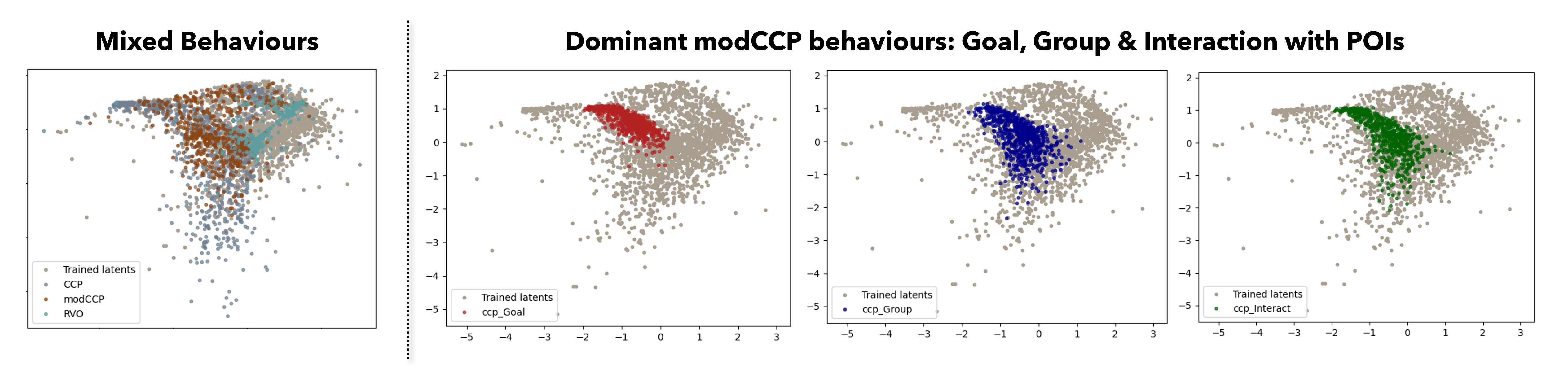}
  \caption[Projections of specific behaviour scenarios onto the embedding space.]{\label{fig:trajeform_coverage} Projections of specific behaviour scenarios onto the embedding space from the principal profiles in modCCP (right), compared to mixture of behaviours predicted for the real data (left).}
\end{figure*}

For instance, inputting only goal, group and interaction-specific behaviours from modCCP weights (e.g., $\{w_{g}=1, w_{gr}=0, w_{i}=0\}$), as shown in Figure~\ref{fig:trajeform_coverage}, shows how well the simulator's principal behaviours can cover the real behaviour space. 
We remind that this experiment references modCCP and not MPACT, since the profiles used are controlled (principal behaviours), and not optimised by the MPACT model.
From Figure~\ref{fig:trajeform_coverage}, it can be seen that the grouping and interaction behaviours have major overlap in the embedding space, which is expected since the dynamics are similar, with the main difference being in the environment context (proximity to POIs). We observe then that these principal behaviours alone, do not span the range of the real behaviours, and so a mixture of them is necessary for obtaining a dataset that can be used, for example, for training models on synthetic data. 

\begin{figure}[h]
  \centering
  \includegraphics[width=\linewidth]{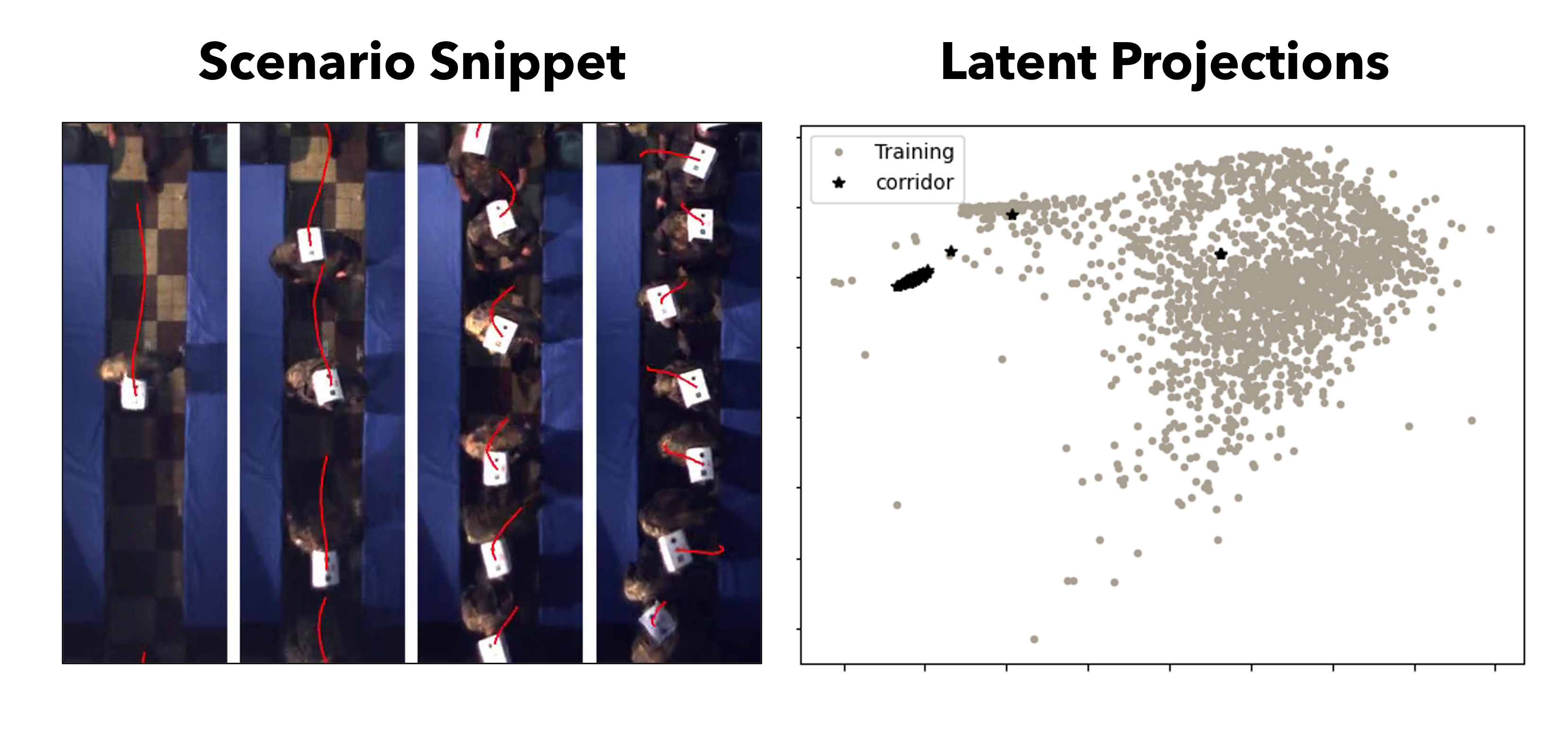}
  \caption[Latents of single file movements.]{\label{fig:corridor} Latents of a controlled scenario: single file movements.}
\end{figure}

Another functionality of our embedding is testing the data projected far from the learned distribution. It is interesting to consider trajectories from a controlled environment, e.g., people instructed to follow a single line~\cite{ped,corridorData}, which is rare to find in in-the-wild data. Figure~\ref{fig:corridor} shows that indeed the embeddings are outside of the learned MoG, and also shows that they are clustered together which is expected since this corridor scenario ``forces'' behaviours to be very similar. 

\begin{figure}[h]
  \centering
  \includegraphics[width=\linewidth]{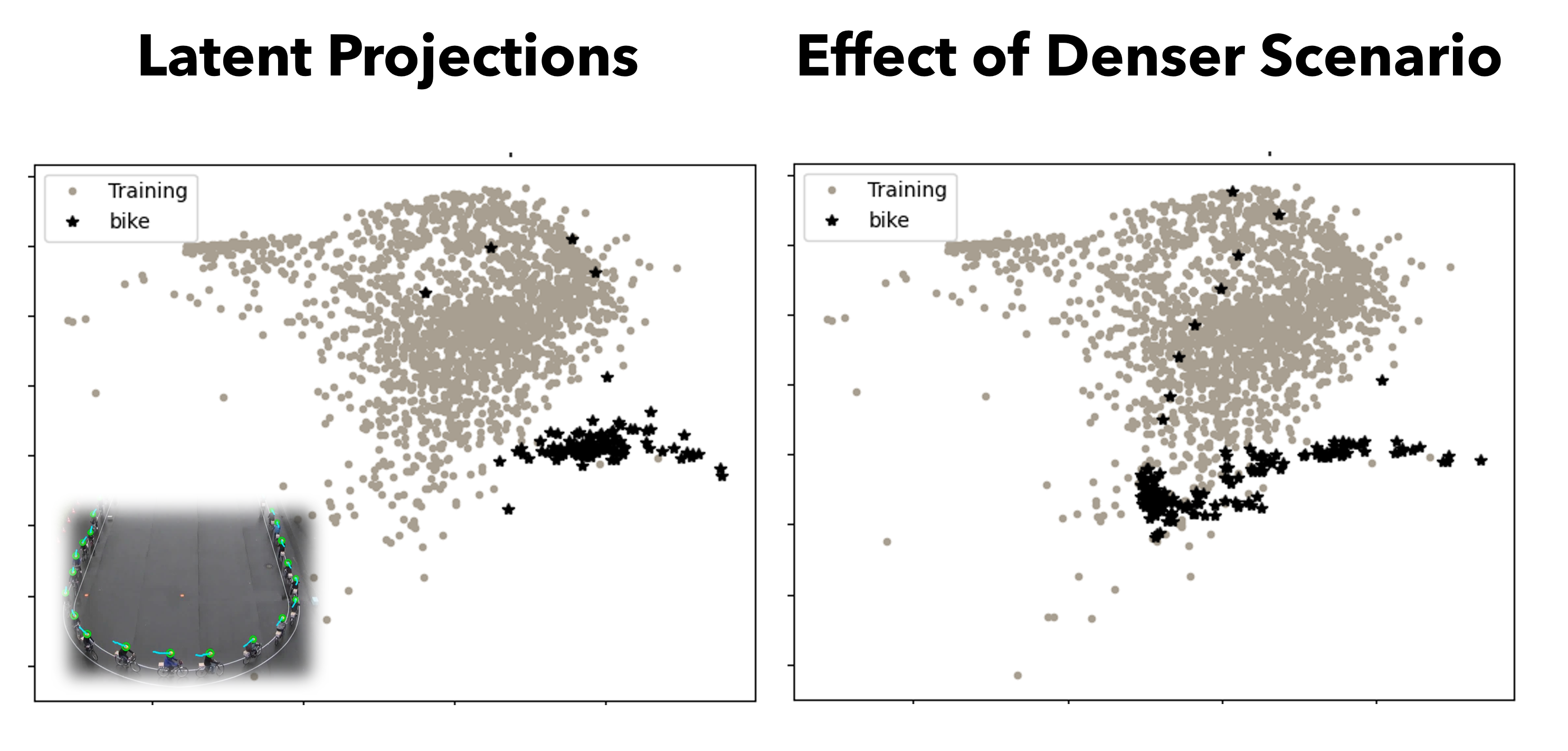}
  \caption[Latents of cyclists' movements in a controlled oval scenario.]{\label{fig:bikes} Latents of cyclists' movements in a controlled oval scenario.}
\end{figure}

\newpage
An additional example of embeddings projected outside the MoG is depicted in Figure~\ref{fig:bikes}, where bike trajectories were captured in a constrained environment~\cite{croma2, bikeData} (data from a project funded by the German Science Foundation (DFG) under DFG-Grant No. KL 1873/1-1 and SE 1789/1-1.). We can also observe the effect of increasing the amount of cyclists in this confined setup, revealing in both instances that the inferred latents fall outside the learned distribution, as expected; not only this oval pattern is rare in natural movements, but the speed differences also affect the perceived behaviours. For additional results, refer to the supplementary material.

\subsection{Guiding the Generations}\label{subs:guidance_Trajeform}

\begin{figure}[t]
  \centering
  \includegraphics[width=\linewidth]{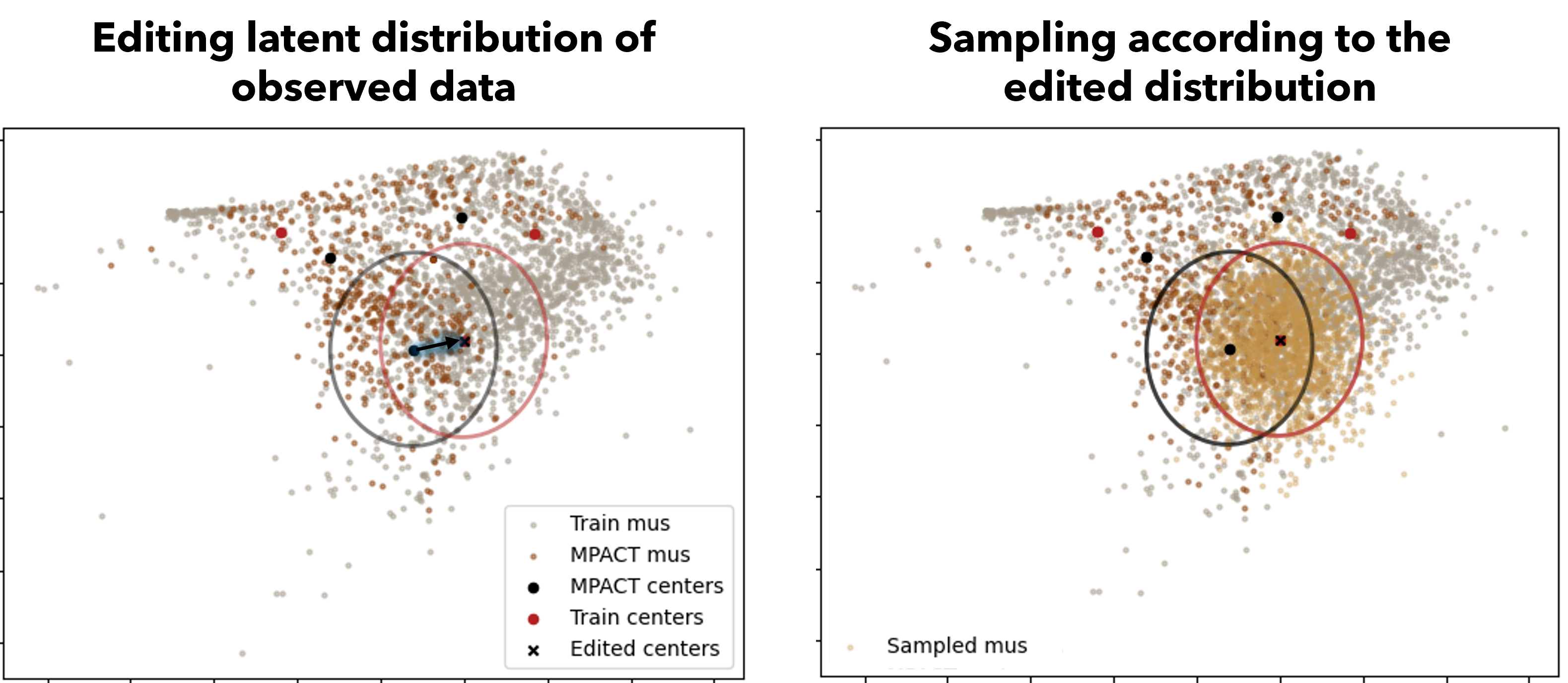}
  \caption[Editing of latent distribution.]{\label{fig:editing} Editing the latent distribution of the observed data. Example of bringing one of its means to coincide with that of the real data (left), and showing the points sampled according to the edited cluster centre (right - yellow points).}
\end{figure}

Provided the posterior distribution of the real data, and a corresponding empirical distribution based on data from a specific simulator, e.g., MPACT~\cite{MPACT}, we can further explore the functionalities of our system by editing the observed distribution. 
Figure~\ref{fig:editing} shows an example of editing the MPACT distribution, specifically shifting one of its inferred cluster centres to coincide with that of the real data. Since the PCA operation is reversible, we can inspect the latent space visualisation (in 2D) and edit accordingly. Then, the new informed distribution incorporates the new cluster mean, and so we can sample from it (with the unedited cluster's frequency and covariance), as shown in Figure~\ref{fig:editing}-right. The latent space ``trajeform'' metric, decreases from $1.867$ (as reported in Table~\ref{table:trajeform_eval2}) to $1.254$, implying that visually editing the  observed latents to have bigger overlap with the trained latents, translates into higher distribution similarity with the real data.

\begin{figure}[h]
  \centering
  \includegraphics[width=\linewidth]{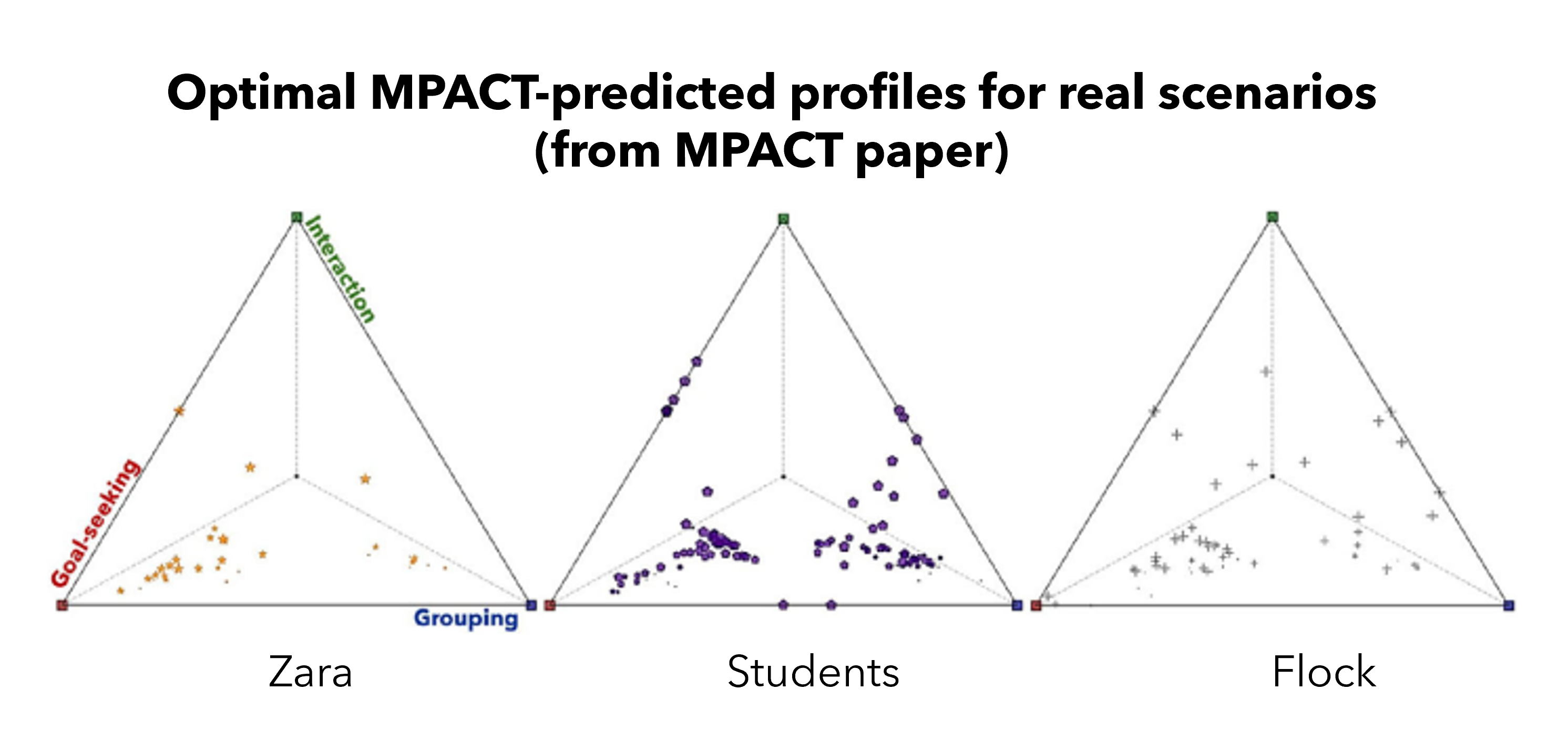}
  \caption[MPACT Space Extract.]{\label{fig:guidance_1.1}  Original projections from MPACT~\cite{MPACT}.}
\end{figure}

\begin{figure*}[h]
  \centering
  \includegraphics[width=.95\linewidth]{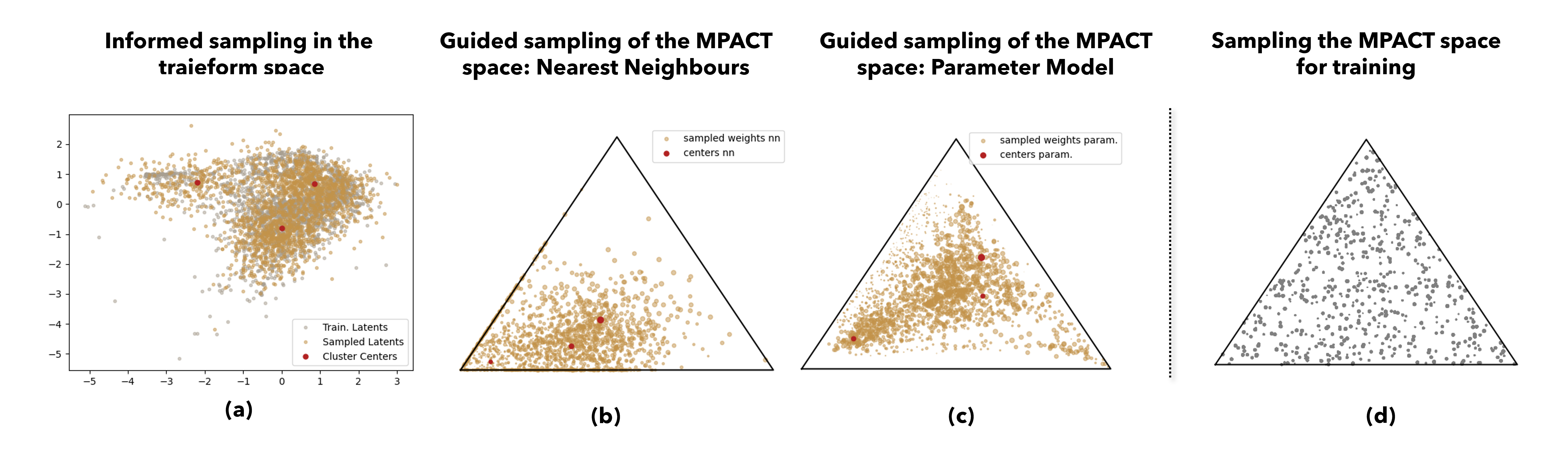}
  \caption[Guiding the generations.]{\label{fig:guidance_1} Generating simulator-specific parameters guided from the informed distribution (a), either via nearest neighbours (b), or a trained parameter model (c). The new set of parameters span the simulator (MPACT) space more faithfully to the real projections compared to sampling the entire space (d).}
\end{figure*}

Since operating on the learned space yields meaningful outcomes, we can then use our model to ``guide'' the selection of parameters in a different parameter space, say $p \in \mathcal{P}$, to create a parameter set that is balanced subject to the real data profiles. 
For this use case, we set $\mathcal{P} \in \mathbb{R}^{4}$ to be modCCP profile space. We remind that modCCP is a simulator that has been used to create the synthetic data to train MPACT - a parameter prediction model. To collect the training data, one could simply randomly sample the parameter space (Figure~\ref{fig:guidance_1}-d). However, the real data do not often span the whole space as can be seen from Figure~\ref{fig:guidance_1.1}-taken directly from the original MPACT paper~\cite{MPACT}. Note here, that for our visualisations we use the same format as MPACT, for uniformity, i.e., converting the goal, group and interaction weights (which sum up to 1) to barycentric coordinates, and denote the connectivity weight (from 0 to 1) via marker size. 

Now, with our model, we can sample our posterior distribution as in Figure~\ref{fig:guidance_1}-a. Then, we can estimate a mapping $\mathcal{M}: \mathcal{Z} \to \mathcal{P}$. One way to do this is via nearest neighbour search: since we have the ground truth weights of the modCCP trajectories passed through our model, we can find the cluster centres in $\mathcal{P}$, and tune accordingly, as in Figure~\ref{fig:guidance_1}-b. We could also learn a simple MLP (Multi-Layer Perceptron) parameter model to predict this mapping, and tune as in Figure~\ref{fig:guidance_1}-c. The parameter model is trained on mean square error loss, on data created by passing modCCP generation (with known weights) through our encoder, and getting the corresponding latents in $\mathcal{Z}$; the model yields training and validation losses $0.0484$ and $0.0499$, respectively, after 200 epochs. We can see that in both cases the tuned points adhere by the real data coverage of $\mathcal{P}$, more than the random sampling. Still, it is important to note a few important considerations. The plotted points in Figure~\ref{fig:guidance_1.1}, taken from the original paper, correspond to profiles per area and per frame-window. In our case, the points correspond to the trajectories, as these are preprocessed by our block, hence there are some discrepancies between the two patterns. A user can trade-off the limited predictive capability of the MLP parameter model with the reliability of nearest neighbours, according to the intended application. 
\section{Evaluation}\label{sec:trajeform_eval}

\subsection{Self-Consistency Check}\label{subs:consistency_trajeform}
\begin{figure}[h]
  \centering
  \includegraphics[width=\linewidth]{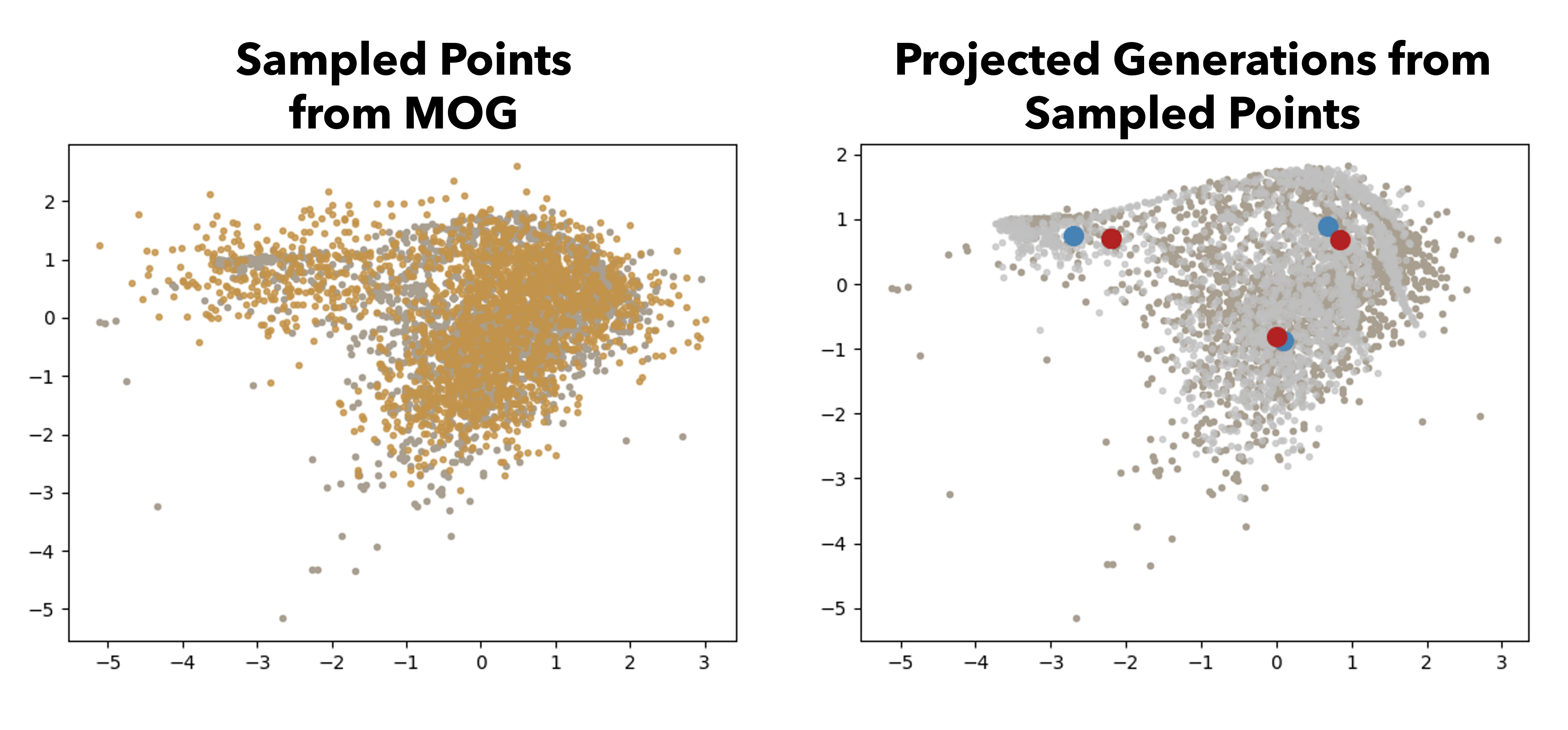}
  \caption[Embedding distribution self-consistency check.]{\label{fig:post_trajeform} Embedding Distribution Self-Consistency Check. Sampled points from the inferred distribution (MoG), generate the corresponding trajectories which are then passed through the encoder to produce new latents, projected in a new posterior distribution which should highly overlap the inferred MoG.}
\end{figure}

Prior to the qualitative and quantitative evaluations, we perform a sanity-check experiment to verify the robustness of the learned embedding space. More specifically, we assess whether sampling latent vectors according to the learned MoG distribution, decoding them into trajectory space, and subsequently re-projecting the generated trajectories back into the latent space through the encoder, results in latent points that still adhere to the learned MoG distribution (Figure~\ref{fig:post_trajeform}). For sampling, we use 2k points to match the size of the training dataset. From Figure~\ref{fig:post_trajeform}, we confirm both visually and quantitatively, through the obtained metric value of $.705$ ($<1$), that the learned latent space is self-consistent.

\subsection{Assessment Premise}\label{subs:premise_traj}

Our aim in this evaluation is to investigate two claims, (i) that following the guidance of our learned latent distribution gives an advantage over no guidance for generation, and (ii) our model encodes information tailored to the real data. Hence, in order to determine the evaluation framework, we discuss each claim and reach a comparison objective that will provide evidence for our claims.

For (i), we use an existing set of trajectories, $t_{h}$, say from simulator, $h$. We compute $z_{R}, z_{h}$ via $z = E_\theta (\tau)$, and sample $\mathcal{Z}$ to get $z_{G}$, where the distribution of our posterior (G) is closer to the real latents than the distribution of $z_{h}$ (e.g., just the sampled prior). Then we use the trained decoder $D_{\theta}: \mathcal{Z} \to \mathcal{T}$ to get decoded $\tau_{G,S}$, where $\tau_{S}$ the decoded version of $\tau_{h}$. Finally we compare $\tau_{G}$ and $\tau_{S}$, aiming to prove that the trajectories with latents more faithful to the distribution of $z_{R}$, are closer to the real trajectories, than the other latents. We use both CCP baseline and the prior as the sampled distributions, and so, we compare the generated trajectories from our \textbf{posterior}, the \textbf{prior}, and the \textbf{CCP} empirical distribution. 

For (ii), let there be $E'$ an alternative encoder, e.g., MPACT, then, $p_{R}=E'(\tau_{R})$. We sample from the distribution of $p_{R}$ to get $p_{G}$. We obtain $\tau'_{G}$ via a known decoder $D'$. We compare our obtained $\tau_{G}$ and $\tau'_{G}$ to the real data. In practice this essentially means comparing the trajectories decoded from two latent distributions: our inferred MoG ($\tau_{G}$), and the distribution of the observed embeddings of another parameter space, \textbf{MPACT} ($\tau'_{G}$). 

Thus, we compare our posterior with MPACT (ii), prior and other simulators (i). We emphasise that all comparisons are made between trajectories generated through our decoder, to ensure that we do not inherently compare simulator specifics, but the embeddings of these simulated data against the real.
\begin{figure*}[t]
  \centering
  \includegraphics[width=.95\linewidth]{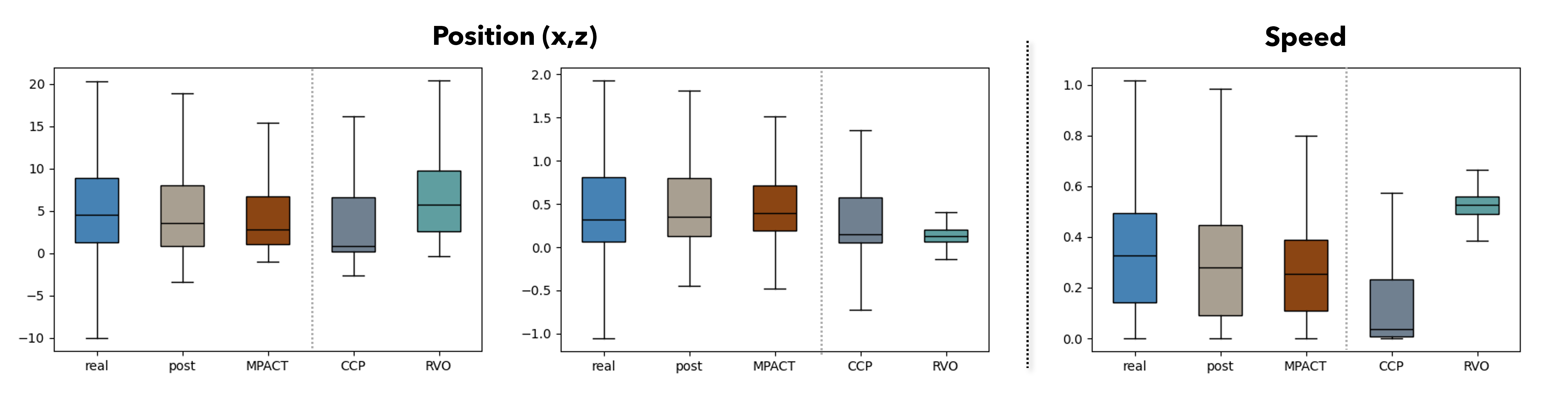}
  \caption[Boxplots of trajectory feature distributions.]{\label{fig:qual_traj} Boxplots of trajectory feature distributions, $(x,z)$ positions and speed, for real data, and decoded data from our posterior samples, MPACT, RVO, and CCP baseline samples.}
\end{figure*}
\begin{figure*}[t]
  \centering
  \includegraphics[width=.95\linewidth]{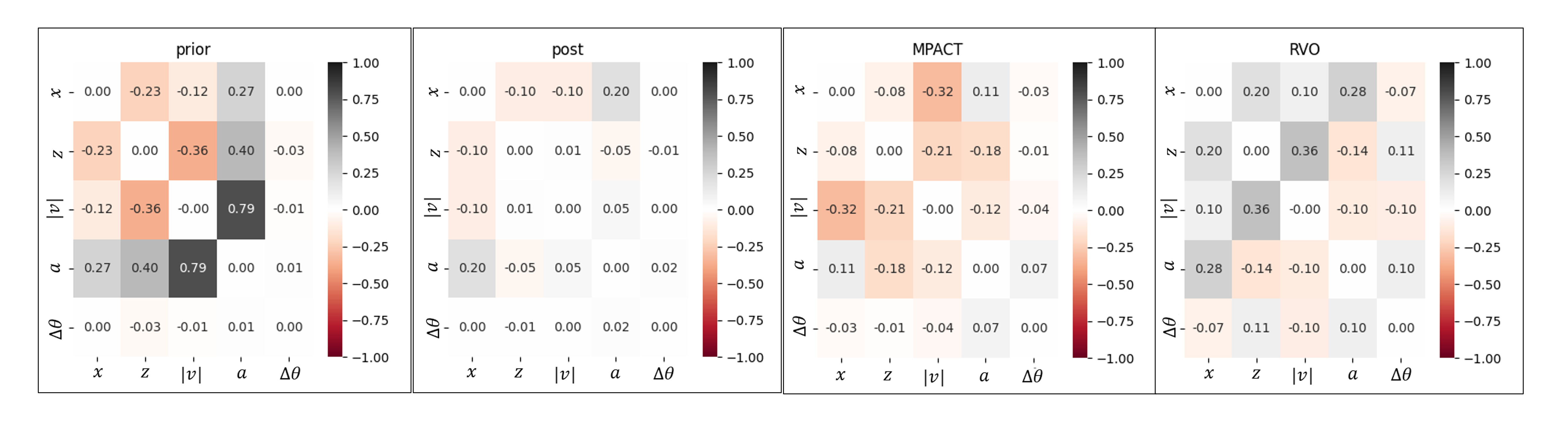}
  \caption[Difference correlation matrices of real against generated data, sampled from different empirical distributions.]{\label{fig:eval_traj} Difference correlation matrices of real against generated data, sampled from different empirical distributions, i.e.,our latent posterior, prior, and empirical RVO and MPACT.}
\end{figure*}

\subsection{Quantitative Assessment}\label{subs:quant_traj}
In order to quantify ``closeness'' to the real data, we sample the latent distributions of five sources, i.e., prior, inferred MoG, RVO, CCP baseline, and MPACT. We compute three types of metrics on trajectories from our decoder for comparison fairness; we want to exclude accumulated limitations of different simulators. So, for each source, we sample their distributions to obtain a set of trajectories and compare them with the real ground truths. Specifically, we calculate: (a) our latent metric described in Section~\ref{subs:generations}, (b) the Maximum Mean Discrepancy-MMD~\cite{gretton2012kernel} values (statistical measure for comparing distributions), 
(c) Sliced Wasserstein Distance~\cite{slicedWD} (SWD), and (d) the energy distance of each set of generated data against the real data.
We use SWD because we do not have 1-1 matching of trajectories and want to measure them containing similar behaviours. SWD measures the similarity between two distributions by comparing their projections along multiple random 1D directions and averaging the resulting Wasserstein distances. The energy distance is a direct distributional distance between two unpaired sets, like MMD, but does not require choosing a kernel.
We finally calculate (e) the Jensen-Shannon Divergence-JSD (symmetric and stable measure for comparing probability distributions): for trajectory features ($x/z$ positions and speeds) and other additional trajectory-based behaviour measures (Deviation From Goal-DFG and Movement Direction Diversity-MDD). 
We report these values in Table~\ref{table:trajeform_eval2}.
\begin{table}[t]
\centering
\begin{tabular}{c|c c: c c c} 
 \hline
 Metric & Prior & MoG & RVO & CCP & MPACT \\ [0.5ex]
 \hline
 Latent Metric$\downarrow$ & 3.833 & \textbf{.705} & 2.732 & 4.433 & \textbf{1.867} \\
 MMD$\downarrow$           & .344  & \textbf{.008} & .030  & .101  & \textbf{.029} \\
 SWD$\downarrow$ & 2.148 & \textbf{.328} & 1.035 & 1.215 & \textbf{.505}\\
 Energy$\downarrow$ & 27.736 & \textbf{.541} & 6.498 & 5.762 & \textbf{1.585}\\
 \hline
 JSD $x\downarrow$         & .442  & \textbf{.007} & \textbf{.014} & .086  & .024 \\
 JSD $z\downarrow$         & .261  & \textbf{.011} & .116  & \textbf{.021} & .032 \\
 JSD speed$\downarrow$     & .422  & \textbf{.004} & .246  & .109  & \textbf{.037} \\
 JSD MDD$\downarrow$       & .103  & \textbf{.005} & .035  & .013  & \textbf{.006} \\
 JSD DFG$\downarrow$       & .443  & .011  & \textbf{.009} & .088  & \textbf{.009} \\
 \hline
\end{tabular}
\caption{
Behaviour Similarity Metrics of Different Posterior-Generated Trajectories. These trajectories are generated from our decoder, subject to latents sampled from a source-specific inferred MoG. We compare these sets of data against the real data, testing for similarity between the distributions of several metrics.}
\label{table:trajeform_eval2}
\end{table}
From Table~\ref{table:trajeform_eval2} we can see that the MoG distribution is closest to the real data in the majority of the metrics, outperforming the prior and MPACT, and so providing evidence for our claims (i) and (ii). Additionally, MPACT achieves overall better scores among the simulator sources, with RVO being second best. We can also see that our metric follows the emergent ``trends'' of the other established metrics, i.e, the ranking of the ``best'' sources is consistent.
The energy distance ranks CCP second best, which again reveals that different metrics focus on different aspects.
Hence, some discrepancies between metrics appear, which is sensible since different metrics often assess different aspects of behaviours/simulations. Therefore, the goal is not to replace existing metrics, but to introduce an additional measure that captures complementary aspects of behavioural similarity.

\begin{table}[b]
\centering
\begin{tabular}{c|c c c } 
 \hline
 Metric & RVO & CCP & MPACT\\ 
 \hline
 Latent Metric$\downarrow$  & 2.732 & 4.433 & \textbf{1.867}\\
 \hdashline
 Latent Energy$\downarrow$ & .891 & .987 & \textbf{.479}\\ 
 MoG WD$\downarrow$ & 1.354 & 2.259 & 1.300\\
 Fr\'echet Distance$\downarrow$ & 1.757 & 3.845 & \textbf{.855}\\
 \hline
\end{tabular}
\caption{
Additional metrics measured on the latent space for the RVO, CCP, and MPACT-generated data projected and compared against the real latents.}
\label{table:trajeform_evalRebuttal}
\end{table}  

Further contextualising the results, Table~\ref{table:trajeform_evalRebuttal} reports an extra set of metrics that operate on the latent space (apart from our latent metric): latent energy, and MoG Wasserstein Distance. 
Fr\'echet Distance~\cite{frechetDistance} measures the similarity between two trajectories by quantifying the minimum maximum distance required to traverse both paths while preserving their point ordering. 
We refrain from using Fr\'echet Distance on the trajectory space since it becomes extremely slow, and instead, we calculate it on the ``core'' trajectories found from each inference run, and report the ranking in decreasing order: MPACT, RVO, and CCP. 
Latent energy and MoG WD ranking coincide with that of the latent metric. 
Overall, the ranking trends provide additional evidence for the quantitative evaluation.

\subsection{Qualitative Assessment}\label{subsqual_traj}
Next, we visualise some of the aforementioned results, for qualitative inspection. Figure~\ref{fig:eval_traj} shows the heatmaps illustrating the difference in the correlation matrices of real and generated data, for four sets of generated data, each associated with a different empirical latent distribution. Correlation matrices~\cite{hotelling1933analysis} are built by computing trajectory features, i.e., positions x and z, speed, acceleration and angular difference, to find whether correlations and dependencies between features are preserved. Taking the difference from the real data matrix, assess representativeness to the real data features. 
We observe that our posterior returns the smallest differences with MPACT being a close second.
Finally, we plot boxplots of the trajectory features (positions and speed) of the real data, and the four sources (excluding the prior) in Figure~\ref{fig:qual_traj}, showing that our MoG posterior and the MPACT generations are closer to the real data, confirming the quantitative assessment observations.
\section{Discussion}\label{sec:trajeform_discussion}

In this work, we learn an encoding that yields a latent representation, embedding real data features, thus parameterising trajectory information. 
To do this, we utilised trajectories captured in real life to train a VAE on the task of trajectory reconstruction. 
We use the trained model encoder to build a posterior distribution (MoG) based on the observed data (latents of the real data), according to predicted cluster centres (k-means clustering). Then we employ our method for embedding synthetic data from various simulators to measure latent distribution similarity. The feature space is aligned with the real one, enabling comparisons between the two sets of latent distributions. We can then inform the selection of latents, and guide it towards a distribution of choice.
This functionality enables measuring behaviour fidelity in an aligned embedding space which is useful for comparing synthetic data to real, as well as for assessing whether new training datasets contain behavioural features that adhere to those of the real ones.  

\paragraph*{Limitations and Future Work}
Despite the new opportunities that our method offers, it has some limitations.
Our model does not take into account environmental information explicitly, e.g., POIs, in its inputs, which limits our system's ability to assign specific semantic meaning to the latent space, e.g., action labels.
Also, it does not directly receive information regarding neighbouring actions and group behaviours. But, these actions are implicitly entangled into the input trajectories, since it falls upon the simulator to produce paths that mimic the real ones, including the spawns and densities of agents. 
Still, it would be beneficial to further experiment with such aspects in the input representation which, however, would require more data since the task becomes more difficult.
\newpage
Additionally, the amount of real data is limited and so the model cannot capture the full range of behavioural features that could emerge. The absence of a large corpus of tracked trajectories captured in-the-wild affects both the encoding power and choice of architecture. Experimenting with more advanced model architectures, e.g., transformers, would be interesting, which entails the creation of a larger training database of real trajectories.
Finally, our method is not extensively assessed against a large variety of alternative quantitative and qualitative metrics, an analysis that would reveal both its strengths and weaknesses. However, the lack of baseline state-of-the-art scores, benchmark evaluation metrics and tasks, makes the proper formulation of such a protocol an interesting but challenging direction for future work. 

\bibliographystyle{IEEEtran}
\bibliography{references}

\end{document}